\documentclass[12pt]{iopart}
\usepackage{bm}
\usepackage{hyperref}
\usepackage{framed}
\usepackage{subfigure}
\usepackage{graphicx}
\usepackage{amssymb}

\expandafter\let\csname equation*\endcsname\relax 
\expandafter\let\csname endequation*\endcsname\relax 
\usepackage{amsmath}
\usepackage{color}
\usepackage{iopams}

\def\be{\begin{equation}}
\def\ee{\end{equation}}

\def\i{\textrm{i}}

\def\tbf{\textbf}

\def\line{\vspace{3mm}}

\begin{document}

\title{Superradiant scattering of dispersive fields}
\author{Mauricio Richartz$^{1}$, Angus Prain$^{2,\,3}$, Silke Weinfurtner$^{2,\,3}$ and Stefano Liberati$^{2,\,3}$}
\address{$^1$ Centro de Matem\'atica, Computa\c{c}\~ao e Cogni\c{c}\~ao, Universidade Federal do ABC (UFABC), 09210-170 Santo Andr\'e, SP, Brazil, \\
$^2$ SISSA - International School for Advanced Studies
via Bonomea 265, 34136 Trieste, Italy, \\
$^3$ INFN, Sezione di Trieste}

\eads{\mailto{mauricio.richartz@ufabc.edu.br}, \mailto{aprain@sissa.it}, \mailto{silkiest@gmail.com}, \mailto{liberati@sissa.it}}

\begin{abstract}

Motivated by analogue models of classical and quantum field theory in curved spacetimes and their recent experimental realizations, we consider wave scattering processes of dispersive fields exhibiting two extra scattering channels. In particular, we investigate how standard superradiant scattering processes are affected by subluminal or superluminal modifications of the dispersion relation. We analyze simple 1+1-dimensional toy-models based on fourth-order corrections to the standard second order wave equation and show that low-frequency waves impinging on generic scattering potentials can be amplified during the process. In specific cases, by assuming a simple step potential, we determine quantitatively the deviations in the amplification spectrum that arise due to dispersion, and demonstrate that the amplification can be further enhanced due to the presence of extra scattering channels. We also consider dispersive scattering processes in which the medium where the scattering takes place is moving with respect to the observer and show that superradiance can also be manifest in such situations. 
\end{abstract}
\maketitle

\section{Introduction and motivation}

Experimental realizations of analogue black holes~\cite{unruh,review} and their associated effects have drawn a lot of attention in the past few years~\cite{unruh_2, philbin, rousseaux, bec_bh, hid_jump,  belgiorno, weinfurtner}. Probably the most discussed results have been the first observation of the classical analogue of Hawking radiation in an open channel flow~\cite{weinfurtner} and the still controversial observation of radiation in ultrashort laser pulse filaments~\cite{belgiorno, unruh_comment, belgiorno_reply, ralf_1, angus_stefano}. Even though it remains an open question whether or not real black holes emit Hawking radiation, calculations involving analogue black holes suggest that the emission process is most probably unaffected by transplanckian effects that could, in principle, alter or even exclude the radiation process~\cite{unruh_3, brout, jacobson}.

    Superradiance~\cite{corinne,super1, super2} is another typical phenomenon of black hole physics~\cite{staro1,staro2} which is also manifest in analogue models of gravity~\cite{basak}. In standard scattering processes, the ratio between the reflected and the incident particle number currents~\cite{wald} (i.e.~the reflection coefficient) is smaller than one. This is directly encoded by the fact that the amplitude of the reflected wave is usually smaller than the amplitude of the incident one in non-dispersive normalized scattering processes. However, in some special situations (e.g.~wave scattering in a Kerr black hole spacetime), low-frequency incident waves can be amplified in the scattering process. This amplification effect, known as superradiance, was first discovered by Zel'dovich in the context of electromagnetic waves~\cite{zeldovich}, and later shown to be a more general phenomenon in physics~\cite{super1,super2}, in which classical as well as quantum field excitations can be amplified.

    The main goal of our work is to analyze the robustness of the amplification process for dispersive fields. The dispersive fields considered in this paper exhibit two extra scattering channels, such that one can study the overall robustness of the amplification process for multiple superradiant scattering. Additionally, in some simplified cases we also investigate the specific deviations in the amplification spectrum due to dispersion. The examples discussed in this paper are motivated by analogue models of gravity, where fields exhibiting sub or superluminal dispersion relations arise naturally.
\subsection{Superradiant systems}

Before starting our analysis of superradiance for dispersive fields, it is important to review the basic ingredients that make the phenomenon possible in some simple systems. For example, in a Kerr black hole, incident scalar field modes of frequency $\omega$ and azimuthal number $m \neq 0$ are known to be superradiant for sufficiently low frequencies.  Although the Klein--Gordon equation is very complicated when expressed in standard Boyer-Lindquist coordinates, the radial part of the equation of motion for these modes can be written, after a change of variables, very simply as
\be
\frac{d^2 u}{dr^* {}^2}+V_{\omega, m}(r^*)u=0,
\ee
where $u$ is related to the radial part of the separated scalar field and $r^*$ is a tortoise-like coordinate (which goes to $-\infty$ at the event horizon and reduces to the standard radial coordinate $r$ at spatial infinity). $V_{\omega, m}(r^*)$ is the effective potential and possesses the following asymptotics,
\be \label{bh2}
V_{\omega, m} (r^*) \rightarrow \begin{cases} (\omega-m\Omega_h)^2, & r^*\rightarrow -\infty, \\  \omega^2, & r^*\rightarrow +\infty, \end{cases}
\ee
where $\Omega_h$ is the angular velocity of the black hole at the event horizon. It can be shown that, for $0<\omega<m\Omega_h$, incident modes are superradiantly scattered by the black hole~\cite{misner, staro1}. The corresponding reflection coefficient is given by
\be \label{basic_ref_kerr}
|R|^2=1-\frac{\omega-m\Omega_h}{\omega}|T|^2 >1,
\ee
where $\frac{\omega-m\Omega_h}{\omega}|T|^2$ is the transmission coefficient.
 Two factors are responsible for the occurrence of the effect~\cite{super2}: first, the event horizon behaves as a one-way membrane that allows only ingoing transmitted waves (defined by the group velocity) near the black hole; second, because of the ergoregion of the rotating black hole, low-frequency ingoing waves are associated with an outgoing particle number current at the horizon (the current being proportional to $\omega-m\Omega_h$)~\cite{wald}. 
This outgoing flux at the horizon is compensated by the superradiantly reflected modes in such a way that the total particle number current is conserved during the process.

Another simple system exhibiting superradiance is a massless scalar field $\phi$ with electric charge $e$ and minimally coupled to an electromagnetic potential $A^\mu = (V(x),0)$ in 1+1-dimensions~\cite{corinne}. The evolution of this system is determined by the Klein--Gordon equation, 
\be \label{kg0}
\left(\partial_{\mu} - i e A_\mu \right)\left(\partial^{\mu} - i e A^\mu \right)\phi=0,
\ee
which, after separation of the temporal dependence (i.e.~$\phi = f(x) \exp(-i \omega t)$), reduces to
\be \label{corinne1}
\frac{d^2 f}{d x^2}+(\omega-eV(x))^2 f=0,
\ee
where $\omega$ is the frequency of the mode in question.
If the electromagnetic potential $V(x)$ has the following asymptotic behaviour,
\be \label{corinne2}
V(x)=\begin{cases} 0, & x\rightarrow -\infty, \\ e\Phi_0, & x\rightarrow +\infty, \end{cases}
\ee
one can calculate the relation between the reflection and transmission coefficients for incident waves~\cite{corinne}: 
\be \label{basic_ref}
|R|^2=1-\frac{\omega-e\Phi_0}{\omega}|T|^2.
\ee
Therefore, low frequency ($0<\omega < e\Phi_0$)\footnote{It is important to remark that the electric potential is defined only up to an arbitrary constant $C$. Therefore, if we redefine $V(x)\rightarrow V(x) + C$, the superradiant condition will be given by $0 < \omega - C <  e \Phi_0 $. Throughout this paper, we fix the potential by requiring that $V(-\infty)=C=0$. Consequently, the condition for superradiance, which is based on the smallness of $\omega - C$, is reduced to a condition on the smallness of $\omega$.} right-moving modes originating from $-\infty$ interact with the scattering potential, resulting in transmitted right-moving modes and superradiantly reflected ($|R|^2>1$) left-moving modes. This is basically the Klein paradox~\cite{klein} -- see~\cite{corinne} for a detailed explanation of the relationship between superradiance, the Klein paradox and pair creation. Note that the important ingredients for superradiance are essentially the same as in the rotating black hole case: firstly, there is no left-moving mode at $+\infty$ (this is similar to the boundary condition at the event horizon of a black hole, where no outgoing solutions are allowed); secondly, the form of the electromagnetic potential allows for low-frequency right-moving modes to be associated with left-moving particle number currents at $+\infty$.

\section{Modified dispersion relations}\label{sec2}

In order to study how a modified dispersion relation affects superradiance, we shall first generalize the simple, non-dispersive, 1+1-dimensional model above~\cite{corinne} by including fourth-order terms in~(\ref{kg0}). 
Inspired by the quartic dispersion relation $\Omega^2=k^2\pm k^4/\Lambda^2$, where $\Lambda$ is a dispersive momentum scale and the $\pm$ notates super and subluminal dispersion respectively, we propose the following generalization of~\eqref{kg0},
\be \label{kgmod}
\mp\frac{1}{\Lambda^2} \partial_x ^4 \phi + \left(\partial_{\mu} - i e A_\mu \right)\left(\partial^{\mu} - i e A^\mu \right)\phi=0,
\ee
which can be obtained from the following action for the complex scalar field $\phi$,
\be \label{action1}
S = \int dtdx \left\{ \left[ \left(\partial_\mu - i e A_{\mu} \right) \phi \right] \left[ \left(\partial^\mu + i e A^{\mu} \right) \phi^* \right] \pm \frac{1}{\Lambda^2} \left(\partial_x^2 \phi \right)\left(\partial_x^2 \phi ^* \right) \right\}.
\ee
After separating the temporal dependence in~\eqref{kgmod}, instead of~\eqref{corinne1}, one obtains
\be
\mp\frac{1}{\Lambda^2}f''''+f''+\left(\omega-eV(x)\right)^2f=0, \label{EOM}
\ee
where $f$ represents the field mode with frequency $\omega >0$ and prime denotes derivative with respect to $x$. We choose the effective frequency $\Omega(x)=\omega-eV(x)$ to satisfy asymptotic relations similar to those in~\eqref{bh2} and \eqref{corinne2}, 
\be
\Omega^2(x)=\begin{cases} \omega^2,\quad &\text{for}\quad x\rightarrow-\infty, \\ \left(\omega-e\Phi_0\right)^2, \quad & \text{for}\quad x\rightarrow+\infty, \quad \end{cases}  \label{asymp}
\ee
where $e\Phi_0$ is a positive constant. In the asymptotic regions, the solutions of~\eqref{EOM} are simple exponentials, $\exp\left(\i kx\right)$, whose wavenumbers $k$ satisfy the dispersion relations below,
\begin{align}
\label{free}
&\mp k^4/\Lambda^2-k^2+\omega^2=0,  &x\rightarrow-\infty, \\ 
\label{interacting}
&\mp k^4/\Lambda^2-k^2+(\omega- e\Phi_0)^2=0, &x\rightarrow+\infty.
\end{align}
	
	In the subluminal case (lower sign), real solutions $(\omega,k)$ to the dispersion relation correspond to the intersections of a lemniscate (figure-eight) and a straight line, see FIG.~\ref{simple_sub}. In  the superluminal case (upper sign), real solutions correspond to the intersections of a quartic curve and a straight line, see FIG.~\ref{superlum}.	
In order to relate the asymptotic solutions at $+\infty$ and $-\infty$ without having to solve the differential equation for all values of $x$, one needs a conserved quantity analogous to the Wronskian for second order wave equations. Since our model is non-dissipative, we expect such a quantity to exist \cite{Laing}. In fact, by calculating the $x$-component of the Noether current associated with the symmetry $\phi \rightarrow e^{i \alpha} \phi$, it is possible to show that the expression
\be \label{wrons_ggg}
Z[f]=W_1+W_2\mp \Lambda^2W_3,
\ee
where 
\begin{align} 
W_1 &=f'''^*f-f^*f''', \notag \\
W_2 &=f'^*f''-f''^*f', \label{W123} \\
W_3 &=f'^*f-f^*f', \notag 
\end{align}
generalizes the notion of the Wronskian\footnote{For the standard 2nd order wave equation, $W_1=W_2=0$ and the functional $W_3$ is the conserved quantity commonly referred to as Wronskian. For Klein--Gordon fields, the Wronskian can be interpreted as the particle number current.}
to our dispersive model, i.e.~$dZ/dx = 0$ for any solution $f$ of~\eqref{EOM}. Furthermore, it is convenient to work with the scaled functional $X$ whose action on a function $f$ is defined by 
\be
X[f]=\frac{Z[f]}{2\i\Lambda^2} \label{gen_wronskian}.
\ee
In particular, the action of $X$ on a linear combination of `on shell' plane waves (wavenumbers satisfying the dispersion relation) is simply
\be 
X\left[ \sum_nA_n\text{e}^{ik_nx}\right]=\sum_n\Omega\frac{d\omega}{dk_n} |A_n|^2. \label{action}
\ee
%
We would like to emphasize the simplicity of the algebraic
expression above, which only depends on the amplitudes, effective
frequencies and group velocities of the various scattering channels participating in the scattering process. Throughout the paper, we shall refer to~(\ref{action}) as the particle number current since it generalizes the usual notion of particle number current associated with a complex Klein-Gordon field.

\subsection{Subluminal scattering}

\label{subzeroflow}

\begin{figure}
\begin{center}

\includegraphics[scale=1]{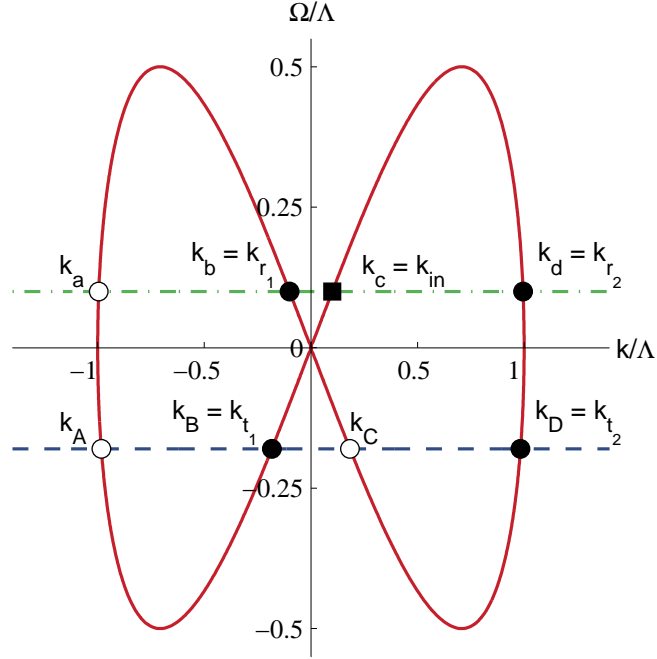}\caption{The dispersion relation for the subluminal case shown in terms of the dimensionless variables $k/\Lambda$ and $\Omega/\Lambda$, where $k$ is the wavenumber and $\Omega=\omega-eV(x)$ is the effective frequency. The green dash-dotted line represents $\Omega=\omega$ when $x\rightarrow-\infty$ while the blue dashed line corresponds to $\Omega=\omega-e\Phi_0$ when $x\rightarrow\infty$. Here $0<\omega<e\Phi_0$ is the fixed lab frequency satisfying the condition for superradiance. We have indicated the incident mode by a black square dot, the transmitted and reflected modes by black circular dots and the remaining solutions of the dispersion relation by white circular dots.
\label{simple_sub}}
\end{center}
\end{figure}

Based on the general notion of superradiance described in the introduction, we will study the scattering process of incident waves originating from $x\rightarrow-\infty$ in the presence of a subluminal dispersion.
In realistic scenarios, we do not expect the dispersion relation $\Omega ^2=k^2 - k^4/\Lambda^2$ to have a maximum/minimum value above/below which only imaginary solutions of the dispersion relation are possible. Therefore, in order to guarantee that at least one real mode is present in the dispersion relation, we assume that the dispersion parameter $\Lambda$ is large compared to the electromagnetic interaction, i.e.~we assume that $e\Phi_0 < \Lambda/2$.     
 In such situations, the behaviour of the solutions to~\eqref{EOM} in the asymptotic limits is captured in FIG.~\ref{simple_sub},
 where the green dash-dotted ($\Omega = \omega$) and the blue dashed ($\Omega = \omega - e \Phi_0$) lines represent the effective frequency when $x\rightarrow-\infty$ and $x\rightarrow+\infty$, respectively. Note that there exist four roots, corresponding to four propagating modes as one would expect from a fourth order differential equation. We also note that, when $\omega<e\Phi_0$, the blue dashed line is located below the $\Omega=0$ axis and the roots in the region $x\rightarrow+\infty$ inherit a relative sign between their group and phase velocities with respect to the roots at $x\rightarrow+\infty$ (compare the intersections of the green dash-dotted line ($x\rightarrow-\infty$) and the blue dashed line ($x\rightarrow+\infty$) with the red solid curve). 

We label the four roots associated with a frequency $0<\omega<e\Phi_0$ as $k_a<k_b<k_c<k_d$ when $x\rightarrow-\infty$ and as $k_A<k_B<k_C<k_D$ when $x\rightarrow+\infty$ (they correspond to the intersections of the straight lines with the figure-eight in FIG.~\ref{simple_sub}). The following table summarizes the character of these modes, where u and v notate right-movers and left-movers respectively,

\vspace{2mm}

\noindent
\begin{tabular}{| p{22mm} | p{3mm} | p{3mm} | p{3mm} | p{3mm} |} 
\hline
Roots at $x\rightarrow-\infty$ &$a$&$b$&$c$&$d$ \\ \hline
group velocity & u& v&u &v  \\ \hline
phase velocity & v&v &u &u\\  \hline
effective frequency & +& +&+ &+  \\ \hline
\end{tabular}
\hspace{0.2mm}
\begin{tabular}{| p{22mm} | p{3.2mm} |p{3.2mm}|p{3.2mm}|p{3.2mm}|} 
\hline
Roots at $x\rightarrow+\infty$ & $A$& $B$& $C$&$D$  \\ \hline
group velocity&v&u&v&u \\ \hline
phase velocity &v&v&u&u\\ \hline
effective frequency &-&-&-&- \\ \hline
\end{tabular}
\vspace{5mm}

Since, in this setup, there is only one source of waves, located at $x\rightarrow-\infty$, and no incoming signal from $x\rightarrow +\infty$, we impose the boundary condition that the modes $A$ and $C$ are unpopulated in the scattering process.
Furthermore, we choose the incoming mode at $x\rightarrow-\infty$ to be entirely composed of low-momentum $c$ modes with no high-momentum $a$ mode component.\footnote{Note that here, and everywhere else in this paper, we consider only stimulated scattering processes in which the ingoing high momentum channels are suppressed. This is not applicable to spontaneous scattering processes, where the quantum vacuum naturally supplies all ingoing, low and high, momentum modes.}
 
 The corresponding solution of~\eqref{EOM} in the asymptotic limits is
\be \label{sol_model}
f \rightarrow \begin{cases} \text{e}^{ik_{in} x} + R_1\text{e}^{ik_{r_1} x}+R_2\text{e}^{ik_{r_2} x}, & x\rightarrow -\infty, \\  T_1\text{e}^{ik_{t_1} x}+T_2\text{e}^{ik_{t_2} x}, & x\rightarrow +\infty, \end{cases}
\ee
where the wavenumber of the incident mode is given by $k_{in} = k_c$, the reflected modes are given by $k_{r_1} = k_b$, $k_{r_2} = k_d$, and the transmitted modes are $k_{t_1} = k_B$, $k_{t_2} = k_D$. In addition, the  coefficients $R_1$, $R_2$, $T_1$ and $T_2$ can be related to reflection and transmission coefficients (see~\eqref{tot_ref} below). 

In the asymptotic regions, it is possible to solve exactly the dispersion relation and find the following explicit expressions for the roots,

\begin{align}
k_{in}&= - k_{r_1}=\frac{\Lambda}{\sqrt{2}}\sqrt{1-\sqrt{1-\frac{4\omega^2}{\Lambda^2}}},  \notag\\
k_{r_2}&=\frac{\Lambda}{\sqrt{2}}\sqrt{1+\sqrt{1-\frac{4\omega^2}{\Lambda^2}}}, \label{momenta} \\
k_{t_1}&=-\frac{\Lambda}{\sqrt{2}}\sqrt{1-\sqrt{1-\frac{4(\omega-e\Phi_0)^2}{\Lambda^2}}},\notag\\
k_{t_2}&=\frac{\Lambda}{\sqrt{2}}\sqrt{1+\sqrt{1-\frac{4(\omega-e\Phi_0)^2}{\Lambda^2}}}. \notag 
\end{align}

In order to compare the particle number current of the reflected waves with the incident and transmitted currents, we substitute~\eqref{sol_model} into the expression for the conserved generalized Wronskian~\eqref{gen_wronskian} and find, after straightforward algebraic manipulation, the following relation between the coefficients $R_1$, $R_2$, $T_1$ and $T_2$,
\be \label{tot_ref}
|R_1|^2+\left|\frac{k_{r_2}}{k_{in}}\right||R_2|^2=1+\sqrt{\frac{\Lambda^2-4(\omega-e\Phi_0)^2}{\Lambda^2-4\omega^2}}\left(\left|\frac{k_{t_1}}{k_{in}}\right||T_1|^2+\left|\frac{k_{t_2}}{k_{in}}\right||T_2|^2\right)>1.
\ee

This relation should be compared to the standard result for non-dispersive 1D scattering, $|R|^2=1-|T|^2$, and its generalization in the presence of an external potential, $|R|^2=1-(\omega-e\Phi_0)|T|^2/\omega$, see~\eqref{basic_ref}. As discussed above, in the non-dispersive case it is possible to achieve $|R|>1$ for sufficiently low frequency scattering with $0<\omega<e\Phi_0$. For fields with subluminal dispersion relations, the conclusion is similar. From expression~\eqref{tot_ref} above, which is valid only when $0 < \omega < e\Phi_0 < \Lambda/2$, we conclude that the total reflection coefficient (i.e.~the ratio between the total reflected particle number current and the incident current) is given by the LHS of~\eqref{tot_ref} and is always greater than one, characterizing a generalization of the usual superradiant scattering which involves extra scattering channels. Note that, in the general case, without an exact solution we have no information about how this total reflection is distributed between the low and high wavenumber channels represented, respectively, by $\vert R_1 \vert ^2$ and $\vert k_{r_2} / k_{in} \vert \vert R_2 \vert ^2$. However, by looking at the $\Lambda$ series expansion of the reflected and transmitted wavenumbers, we can draw some interesting conclusions about the regime $\Lambda\gg 1$. 
Using~\eqref{momenta}, we obtain the following expansions for the relevant wavenumbers,
\begin{align}
k_{in}&= - k_{r_1}=\omega + \frac{\omega^3}{2 \Lambda ^2} + \mathcal{O}(\Lambda^{-3}),  \notag\\
k_{r_2} &= \Lambda - \frac{\omega^2}{2 \Lambda}  + \mathcal{O}\left(\Lambda^{-2}\right), \label{momenta_exp} \\
k_{t_1}&= \omega - e\Phi_0 + \frac{\left( \omega - e\Phi_0 \right)^3}{2 \Lambda ^2} + \mathcal{O}(\Lambda^{-3}),\notag\\
k_{t_2}&= \Lambda - \frac{\left( \omega - e\Phi_0 \right)^2}{2 \Lambda}  + \mathcal{O}\left(\Lambda^{-2}\right). \notag 
\end{align}

\begin{figure}
\begin{center}
\includegraphics[scale=1.0]{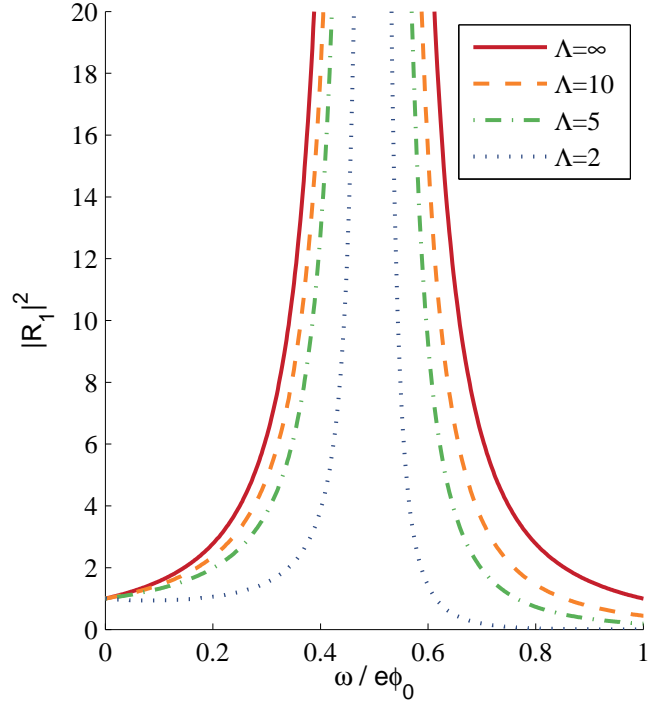}\caption{The reflection coefficient $|R_1|^2$ for the subluminal case as a function of $\omega/e\Phi_0$ calculated for the step potential $V(x)=e\Phi_0\Theta(x)$ with $e\Phi_0=1$. Note the singular behaviour at $\omega = e\Phi_0 / 2$, which is present even in the non-dispersive case. This divergence is directly related to the discontinuity in $V(x)$ and would be cured by smoothing the step potential at $x=0$. \label{step1}}
\end{center}
\end{figure}

\begin{figure}
\begin{center}
\includegraphics[scale=1.0]{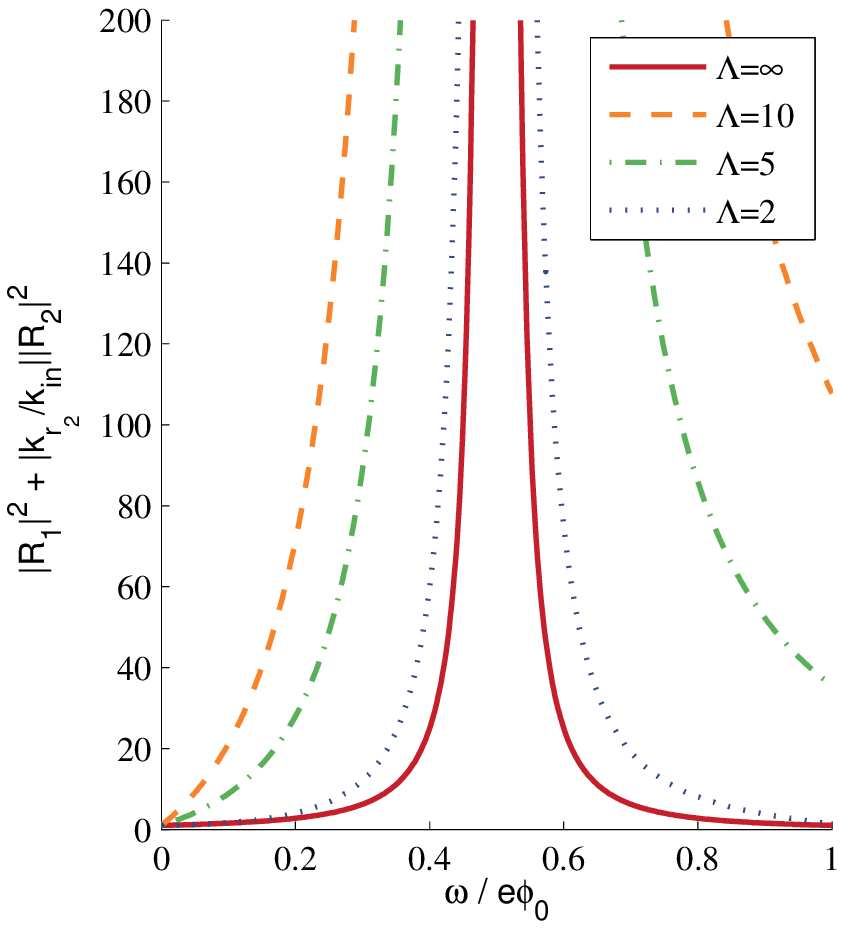}\caption{The total reflection coefficient $\vert R_1 \vert ^2 + \vert k_{r_2} / k_{in} \vert \vert R_2 \vert ^2$ as a function of $\omega/e\Phi_0$ for a subluminal dispersion relation. Calculations were performed assuming a step potential $V(x)=e\Phi_0\Theta(x)$ with $e\Phi_0=1$. Note the enhancement effect in the amplification caused by dispersion.  \label{step11}}
\end{center}
\end{figure}

As mentioned before, all possible modes of the system are characterized by the same conserved frequency $\omega$. The momentum of each mode, on the other hand, is determined by the wavenumber $k$ and, therefore, is not the same for every mode.
From the expansions above, we note that $k_{t_{2}}, k_{r_2} \sim \mathcal{O}(\Lambda)$ while $k_{in},k_{r_1},k_{t_1} \sim \mathcal{O}(\Lambda^0)$. In particular, the difference between the high momentum modes $k_{r_2}$ and $k_{t_2}$ is $k_{r_2} - k_{t_2} \sim \mathcal{O}\left(\Lambda^{-1}\right)$. 
If the potential is sufficiently smooth, this suggests that the creation of a pair of these modes $\left(k_{r_2},k_{t_2}\right)$ should be favoured since it requires a negligible momentum change in the system if $\Lambda \gg 1$.

It is also interesting to analyse the case of a non-smooth potential, by solving the idealized problem of a step function, i.e.~$V(x)=e\Phi_0\Theta(x)$ (see the appendix for a discussion concerning the appropriate boundary conditions used at the discontinuity point $x=0$). In such a case, one can show that the reflection and transmission coefficients are given by,
\begin{align}
R_1 = \frac{(k_{in}-k_{r_2}) (k_{in}-k_{t_1}) (k_{in}-k_{t_2})}{(k_{r_2}-k_{r_1}) (k_{r_1}-k_{t_1}) (k_{r_1}-k_{t_2})}, \nonumber \\
R_2= \frac{(k_{in}-k_{r_1}) (k_{in}-k_{t_1}) (k_{in}-k_{t_2})}{(k_{r_1}-k_{r_2}) (k_{r_2}-k_{t_1}) (k_{r_2}-k_{t_2})}, \\
T_1= \frac{(k_{in}-k_{r_1}) (k_{in}-k_{r_2}) (k_{in}-k_{t_2})}{(k_{r_1}-k_{t_1}) (k_{r_2}-k_{t_1}) (k_{t_1}-k_{t_2})}, \nonumber \\
T_2= \frac{(k_{in}-k_{r_1}) (k_{in}-k_{r_2}) (k_{in}-k_{t_1})}{(k_{r_1}-k_{t_2}) (k_{r_2}-k_{t_2}) (k_{t_2}-k_{t_1})}, \nonumber
\end{align} 
where the wavenumbers $k$ are explicitly given in~\eqref{momenta}. Note that the expressions above are valid for any value of $\Lambda$, not only in the regime $\Lambda \gg 1$. In order to observe the effects of the dispersive parameter, we plot the reflection coefficient $|R_1|^2$ as function of $\omega$ for different values of $\Lambda$, see FIG.~\ref{step1}. Note that as we increase the dispersive effects (i.e.~we lower $\Lambda$), the reflection coefficient associated with $k_{r_{1}}$ decreases. We can also see in FIG.~\ref{step1} that the $R_1$ channel even becomes non-superradiant for some frequencies in the range $0<\omega< e\Phi_0$. Of course, if we also include the other reflection channel, see FIG.~\ref{step11}, then the total reflection coefficient is always superradiant, as proven in~\eqref{tot_ref}. Note that the presence of extra scattering channels even enhances the amplification process in comparison with the non-dispersive case. This can be understood in the sense that the amplifier is more effective if the number of `accessible' channels increases.
 
Note also that the total reflection coefficient in the dispersive case is not continuously connected with the non-dispersive regime. More precisely, as $\Lambda$ is increased in FIG.~\ref{step11}, the total reflection coefficient becomes larger and larger and moves away from the non-dispersive coefficient ($\Lambda = \infty$). A possible explanation for this behaviour resides in the fact that the subluminal dispersion relation itself is not continuously connected to the linear dispersion relation in the limit $\Lambda \rightarrow \infty$. In other words, the modes $k_{r_2}$ and $k_{t_2}$ which are always absent in the non-dispersive case, will be present in the dispersive regime no matter how large $\Lambda$ is. 
 
\subsection{Superluminal scattering}

\begin{figure}
\begin{center}
\includegraphics[scale=1.0]{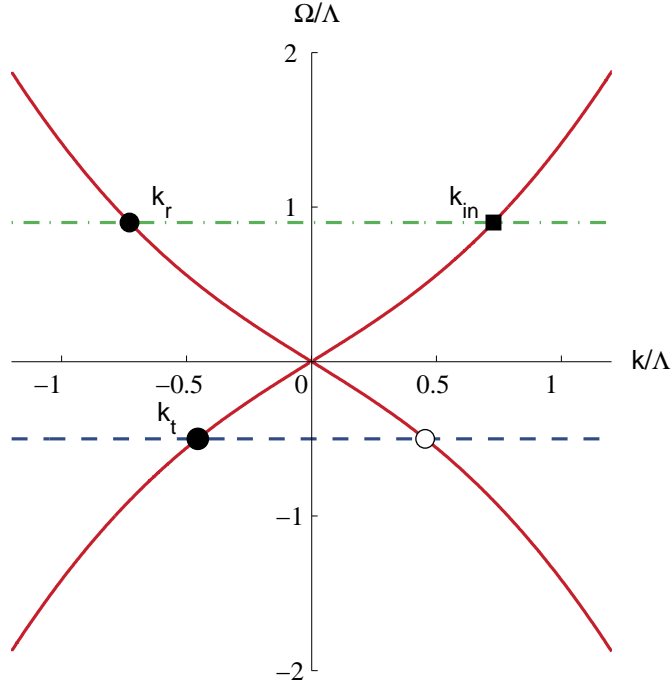}\caption{The dispersion relation for the superluminal case shown in terms of the dimensionless variables $k/\Lambda$ and $\Omega/\Lambda$. We have chosen here a frequency in the superradiant interval $0<\omega<e\Phi_0$. The blue dashed line corresponds to the dispersion relation at $x\rightarrow+\infty$ while the green dash-dotted line corresponds to the dispersion at $x\rightarrow-\infty$. The incident mode is indicated by a black square dot, the transmitted and reflected modes by black circular dots and the remaining solutions of the dispersion relation by white circular dots. \label{superlum}}
\end{center}
\end{figure}

Let us now turn our attention to the case of a superluminal dispersion. Once again, we can understand much of the scattering process from the dispersion relation, see FIG.~\ref{superlum}.
The main difference with the subluminal case is that now there are only two real roots in either asymptotic regions, corresponding to one left-moving and one right-moving mode. The other two solutions to the fourth order equation are imaginary roots, corresponding to exponentially decaying and exponentially growing modes. Similarly to the subluminal case, we are interested in the scattering of an incident low-frequency ($0<\omega < e\Phi_0$) wave originating from $x\rightarrow -\infty$, which is converted into a reflected left-moving mode, a transmitted right-moving mode and exponentially decaying modes (as a boundary condition, we impose the fact that there can be no exponentially growing modes in the asymptotic regions). Therefore, the solution of~\eqref{EOM} corresponding to this scattering process is given by
\be \label{sol_model2}
f \rightarrow \begin{cases} \text{e}^{ik_\text{in}x} +R \text{e}^{ik_rx} + E_r \text{e}^{k_{er}x}, & x\rightarrow -\infty, \\  T \text{e}^{ik_tx} +E_t \text{e}^{-k_{et}x}, & x\rightarrow +\infty, \end{cases}
\ee
where the coefficients $R$ and $T$ are related, respectively, to reflection and transmission coefficients (see~\eqref{wrk2} below), and the coefficients $E_r$ and $E_t$ are the coefficients of the exponentially decaying modes. The wavenumbers $k$ in~\eqref{sol_model2} are obtained directly from the dispersion relations~\eqref{free} and \eqref{interacting},
\begin{align}
k_\text{in}&= -k_r=\frac{\Lambda}{\sqrt{2}}\sqrt{-1+\sqrt{1+\frac{4\omega^2}{\Lambda^2}} },  \notag\\
k_{er}&=\frac{\Lambda}{\sqrt{2}}\sqrt{1+\sqrt{1+\frac{4\omega^2}{\Lambda^2}} }, \label{momenta2}\\
k_t&=-\frac{\Lambda}{\sqrt{2}}\sqrt{-1+\sqrt{1+\frac{4(\omega-e\Phi_0)^2}{\Lambda^2}} }, \notag \\
k_{et}&=\frac{\Lambda}{\sqrt{2}}\sqrt{1+\sqrt{1+\frac{4(\omega-e\Phi_0)^2}{\Lambda^2}} }. \notag
\end{align}
 
The conservation of particle number current can be obtained by inserting~\eqref{sol_model2} into~\eqref{action} and equating the generalized Wronskian at $\pm\infty$. Compared to the non-dispersive result, one might expect extra terms related to the exponentially decaying modes. However, these extra contributions are also exponentially decaying and, therefore, their particle number currents are negligible at $\pm \infty$. As a consequence, one obtains the following reflection coefficient (valid for $0<\omega <e\Phi_0$),
\be \label{wrk2}
|R|^2 = 1 + \sqrt{ \frac{\Lambda^2+4(\omega-e\Phi_0)^2}{\Lambda^2+4\omega^2}}  \left|\frac{k_t}{k_\text{in}}\right| |T|^2>1,
\ee
which, similarly to the subluminal case, is always greater than one. Note also that the expression above reduces to the usual non-dispersive reflection coefficient \eqref{basic_ref} in the limit $\Lambda \rightarrow \infty$.

\begin{figure}[h!]
\begin{center}
\includegraphics[scale=1.0]{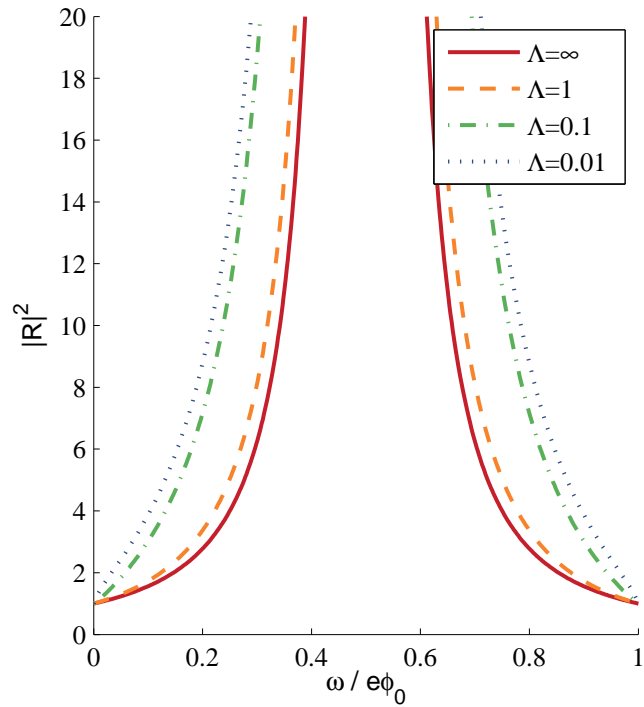}\caption{The reflection coefficient $|R|^2$ for the superluminal case as a function of $\omega/e\Phi_0$ calculated for the step potential $V(x)=e\Phi_0\Theta(x)$ with $e\Phi_0=1$. Like in the subluminal case, the divergence at $\omega = e\Phi_0 / 2$ is caused by the discontinuity in $V(x)$ and would be absent if a smooth potential were used. Note the enhancement in amplification due to dispersion.\label{step2}}
\end{center}
\end{figure}

Repeating the analysis used in the subluminal case, one can solve exactly the problem for a step potential given by $V(x)=e\Phi_0\Theta(x)$. Using appropriate boundary conditions at $x=0$ (see the appendix), one can relate the coefficients in~\eqref{sol_model2} to the wavenumbers given in~\eqref{momenta2},
\begin{align}
R &= \frac{(ik_{in}+k_{et}) (k_{in}+ik_{er}) (k_{in}-k_{t})}{(k_{er}-ik_{r}) (k_{r}-ik_{et}) (k_{r}-k_{t})}, \nonumber \\
E_r &= \frac{(ik_{in}+k_{et}) (k_{in} - k_{r}) (k_{in}-k_{t})}{(k_{er}+ k_{et}) (k_{er}-ik_{r}) (k_{er}-ik_{t})}, \\
T &= \frac{(ik_{in}-k_{er}) (k_{et}+ik_{in}) (k_{in}-k_{r})}{(k_{r}-k_{t}) (k_{t}-ik_{er}) (k_{t}-ik_{et})}, \nonumber\\
E_t &= \frac{(ik_{in}-k_{er})(k_{in}-k_{r}) (k_{in}-k_{t})}{(k_{er}+k_{et}) (k_{et}+ik_{r}) (k_{et}+k_{t})}. \nonumber
\end{align} 
Once again, it is useful to plot the reflection coefficient as a function of $\omega$ for different values of the dispersive parameter $\Lambda$, see FIG.~\ref{step2}. 

The results obtained for the scattering of sub and superluminal dispersive fields in our 1+1-dimensional toy-model demonstrate that superradiance is possible and that the amplification is enhanced due to the extra scattering channels in comparison with non-dispersive superradiance. These results are interesting for experimental attempts of detecting superradiance, since they indicate that dispersion may increase the amplification rates and, consequently, make the effect easier to observe.
\\

\section{Generalization to non-zero flows} \label{sec2c}

Having in mind moving media in analogue models of gravity, we extend the ideas of the previous section by including a position--dependent flow velocity $W(x)$ in our model. In other words, the medium where the scattering process takes place (fluid frame) is in relative motion with respect to the observer (lab frame). If we require that the dispersion relation be unaltered in the comoving frame of the fluid, the 
action~\eqref{action1} generalizes to
\begin{align}
S = \int dtdx \left\{ \left[ \left(\partial_t + W \partial_x - i e A_{t} \right) \phi \right] \left[ \left(\partial^t -W \partial_x + i e A^{t} \right) \phi^* \right] \right. \nonumber \\
\left. + \left(\partial_x \phi \right)\left(\partial^x \phi^* \right) \pm \frac{1}{\Lambda^2} \left(\partial_x^2 \phi \right)\left(\partial_x^2 \phi ^* \right) \right\}.  \label{action2}
\end{align}
Consequently, the modified Klein--Gordon equation~\eqref{kgmod} generalizes to
\be
\mp\frac{1}{\Lambda^2}\partial_x ^4 \phi + \partial_x^2 \phi + \left(\partial_t - ie A_t + \partial_x W \right)\left(\partial ^t -i e A^t - W \partial_x \right)\phi=0 \label{kgmod2},
\ee
%
where the derivative operator $\partial_x$ is understood to act on everything to its right. It is important to remark that the reference velocity $c=1$ in such analogue systems corresponds to the velocity of the scalar perturbations (e.g.~the sound speed in hydrodynamical systems). Usually, such velocities (and also $W(x)$) are much smaller than the speed of light and, therefore, the system is nonrelativistic. For relativistic analogue models of gravity, we refer the reader to \cite{relat1,relat2,relat3}.

 After separation of variables, $\phi(t,x) = e^{-i \omega t} f(x)$, the wave equation~\eqref{kgmod2} becomes,
\be
\mp\frac{1}{\Lambda^2}\frac{d^4f}{dx^4}+\frac{d^2f}{dx^2}+\left(\omega-eV(x)+\i\frac{d}{dx}W(x)\right)\left(\omega-eV(x)+\i W(x)\frac{d}{dx}\right)f=0 \label{E:model},
\ee
%
which can be written as
\be 
f''''(x)+\alpha(x) f''(x)+\beta(x)f'(x)+\gamma(x)f=0, \label{nonzeroflow}
\ee
where
\begin{align}
\alpha &=\pm\Lambda^2\left(1-W^2(x)\right) \notag \\
\beta &=\mp\Lambda^2\left[2W(x)\partial_xW(x)-2\i W(x)\left(\omega-eV(x)\right)\right]  \\
\gamma &=\pm\Lambda^2\left[(\omega-eV(x))^2+\i\partial_x\left(W(x)(\omega-eV(x))\right)\right]. \notag
\end{align}
Observe that these coefficients are not independent but satisfy the following relations,
\begin{align}
&\gamma-\gamma^*=2i\mathcal{I}m(\gamma)= i \mathcal{I}m(\partial_x \beta) =   \partial_x\left(\frac{\beta - \beta^*}{2}\right), \label{wcond1} \\
&\mathcal{R}e(\beta)=\frac{1}{2}\left(\beta+\beta^*\right)=\partial_x\alpha. \label{wcond2} 
\end{align}
We choose the functions $W(x)$ and $V(x)$ to be asymptotically constant,
\begin{align}
W(x)&=\begin{cases} 0, & x\rightarrow -\infty,\\ W_0, & x\rightarrow +\infty, \end{cases}\\
eV(x)&=\begin{cases} 0, & x\rightarrow-\infty, \\ e\Phi_0, & x\rightarrow +\infty, \end{cases}
\end{align}
so that, at $\pm \infty$, any solution to~\eqref{E:model} can be decomposed into plane waves satisfying the dispersion relation below,
\be
k^2\pm\frac{k^4}{\Lambda^2}=\begin{cases} \omega^2, & x\rightarrow -\infty, \\ \left(\omega-e\Phi_0-kW_0\right)^2, & x\rightarrow +\infty, \end{cases} \label{disp_flow}
\ee
where the $\pm$ stands for super and subluminal respectively. As previously, we work under the assumption that $\omega,e\Phi_0\ll\Lambda$.

Similarly to the zero-flow case, we can obtain a conserved quantity by calculating the $x$-component of the Noether current associated with the symmetry $\phi \rightarrow e^{i \alpha} \phi$. After separating the temporal dependence in $\phi$, one can show that a modification of~\eqref{wrons_ggg}, namely
\be \label{nonzero_wronsk}
Z=W_1+W_2+\alpha W_3-\i\;\mathcal{I}m(\beta)|f|^2,
\ee
and the corresponding scaled quantity $X=Z/(2\i\Lambda^2)$ are independent of $x$.   The action of the functional $X$ on a linear combination of `on-shell' plane waves takes precisely the same form as in the zero-flow case, see~\eqref{action} (the only difference is that the effective frequency is now given by $\Omega=\omega-eV-kW$ instead of $\Omega=\omega-eV$). This result highlights the generality of the algebraic structure of the particle number currents given by~\eqref{action}.
 Additionally, the existence of superradiance in the zero-flow case was related to the condition that $\omega-e\Phi_0<0$. We can anticipate that the occurrence of superradiance in non-zero flows will be favoured by modes for which $\omega-e\Phi_0-kW_0<0$.

\subsection{Subluminal dispersion}

\begin{figure}[h!]
\begin{center}
\includegraphics[scale=1.0]{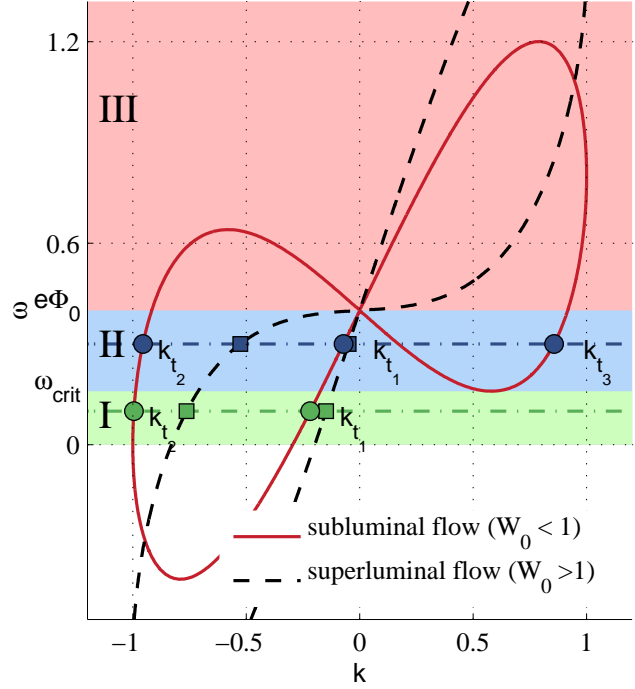}
\caption{The subluminal dispersion curve in the lab frame ($\Lambda=1$, $e\Phi_0=0.4$) at $x\rightarrow +	\infty$ for two different flow velocities: $e\Phi_0/\Lambda<W_0<1$ (red solid curve) and $W_0>1$ (black dashed curve).  The coloured regions described in the text are associated with the red solid curve: region I (green, $0<\omega<\omega_\text{crit}$),  region II (blue, $\omega_\text{crit}<\omega<e\Phi_0$), and region III  (light red, non-superradiantly scattering region). The intersections of the horizontal lines with the red solid curve and the black dashed curve indicate transmitted modes at that frequency. Note that, when $x\rightarrow -	\infty$, the dispersion is described by the green dash-dotted curve of FIG.~\ref{simple_sub}.} \label{regions_lab_sub}
\end{center}
\end{figure}

Let us consider first the case of a subluminal dispersion relation. 
As shown in FIG.~\ref{regions_lab_sub} (red solid curve), for fixed $e\Phi_0/\Lambda <W_0<1$, there are two distinct intervals of frequencies separated by a critical frequency $\omega_\text{crit} < e\Phi_0$ in which we expect superradiance: for $0<\omega<\omega_\text{crit}$ (region I) two propagating modes are admitted, both right-moving in the lab frame;  for $\omega_\text{crit}<\omega<e\Phi_0$ (region II) there are four real roots of the dispersion relation corresponding to four propagating modes, three right-moving and one left-moving (with respect to the lab-frame). Note that the requirement that $W_0$ is not too small, specifically $W_0>e\Phi_0/\Lambda$, is necessary in order to guarantee that the left-most root in region I (i.e.~the circular dot labeled $k_{t_2}$ in region I of FIG.~\ref{regions_lab_sub}) has positive group velocity and hence defines a true transmitted mode. If $0<W_0<e\Phi_0/\Lambda$, the situation is basically the same as the one discussed in section \ref{subzeroflow}. On the other hand, if $W_0 > 1$ (see the black dashed curve in FIG.~\ref{regions_lab_sub}), there is only one possible regime: for all frequencies $0< \omega < e\Phi_0$, two right-moving modes are admitted.  

An interesting fact is that, due to the absence of left-moving modes at $+\infty$ when $\omega < \omega_\text{crit}$, this system is a model for the event horizon of an analogue black hole with modified dispersion relations. However, the precise location of the horizon, besides being $\omega$-dependent, is also rather ill-defined, relying on a global solution to the equation of motion in the vicinity of a classical turning point. This region can be studied by WKB methods and Hamilton--Jacobi theory \cite{coutant}, but it is not of specific interest to us here.

According to the analysis above, the scattering of an incoming wave from $-\infty$ will result in transmission through two or three channels, depending on whether $\omega < \omega_\text{crit}$ or not. An exact solution to the scattering problem can be decomposed as
\be \label{sol_model3}
f \rightarrow \begin{cases} \text{e}^{ik_{in} x} + R_1\text{e}^{ik_{r1} x}+R_2\text{e}^{ik_{r2} x}, & x\rightarrow -\infty, \\  T_1\text{e}^{ik_{t1} x}+T_2\text{e}^{ik_{t2} x}, & x\rightarrow +\infty, \end{cases}
\ee
when $0 < \omega < \omega_\text{crit}$ (if $W_0 < 1$) or $0 < \omega < e\Phi_0$ (if $W_0>1$) and as
\be \label{sol_model3b}
f \rightarrow \begin{cases} \text{e}^{ik_{in} x} + R_1\text{e}^{ik_{r1} x}+R_2\text{e}^{ik_{r2} x}, & x\rightarrow -\infty, \\  T_1\text{e}^{ik_{t1} x}+T_2\text{e}^{ik_{t2} x}+T_3\text{e}^{ik_{t3} x}, & x\rightarrow +\infty, \end{cases}
\ee
when $ \omega_\text{crit} < \omega < e\Phi_0$ (only possible if $W_0<1$). The wavenumbers $k_{in}$, $k_{r_1}$ and $k_{r_2}$ are given by~\eqref{momenta} and are labeled in FIG.~\ref{simple_sub}, while the transmitted wavenumbers are not, in general, expressible as simple functions of the parameters.  Note that we do not explicitly keep track of possible exponential decaying solutions in~\eqref{sol_model3} (corresponding to complex solutions to the dispersion relation), since they do not contribute directly to~\eqref{wrons_subl}.

We insert the solutions \eqref{sol_model3} and \eqref{sol_model3b} into the functional $X$ of~\eqref{action} and find, after algebraic manipulations, the following relationship between the transmission and reflection coefficients,
\be \label{wrons_subl}
|R_1|^2+\left|\frac{k_{r_2}}{k_{in}}\right||R_2|^2 = 1 - \frac{\Lambda}{\sqrt{\Lambda ^2 - 4 \omega ^2}} \left(\sum_n \frac{v_{g_n}}{k_{in}} \left( \omega - e\Phi_0 - k_{t_n} W_0 \right)  \left| T_n \right|^2   \right),
\ee
where the sum is over all (2 or 3, depending on $\omega$ and $W_0$) transmission channels. Here, $v_{g_n}=v_g(k_{t_n})$ are the group velocities of the transmitted modes at $+\infty$. The LHS of~\eqref{wrons_subl} can be interpreted as the total reflection coefficient associated with the incident modes. 
 
Since the group velocities $v_{g_n}$ of the transmitted modes are, by definition, always positive, the sign of the contribution from each channel $k_{t_{n}}$ to the RHS of \eqref{wrons_subl} is determined by the factor $\Omega(k)=\omega-e\Phi_0-kW_0$ evaluated at $k_{t_n}$, as we anticipated previously. When only two transmission channels are admitted, one of the two factors $\Omega(k)$ is strictly negative while in the case of three transmission channels, two of the three factors $\Omega(k)$ are strictly negative. Therefore, in both situations there is one root which contributes an overall \textit{negative} amount to the RHS of~\eqref{wrons_subl} and thus reduces the magnitude of the total reflection. Because of these troublesome modes, the RHS is not strictly greater than $1$ and we cannot straightforwardly conclude superradiance. In general, in order to fully answer the question of superradiance, one would need to specify $W(x)$ and $V(x)$ in all space and solve for the coefficients $R_n$ and $T_n$. 

\line 

\noindent\tbf{Large $\Lambda$ approximation:} To better understand the relation between~\eqref{wrons_subl} and superradiance, we now focus on small deviations ($\Lambda \gg 1$) from the non-dispersive case. For fixed $W_0<1$,
 the size of region I in FIG.~\ref{regions_lab_sub} becomes zero when  $\Lambda$ is greater then $e\Phi_0\left[2(1-W_0)/3\right]^{-2/3}$. We therefore assume that the frequency $\omega$ lies in region II, where three transmission and two reflection channels are available, see~\eqref{sol_model3b}. As explained above, in the most general case one would need to solve the equation of motion for all $x$ in order to conclude superradiance or not from~\eqref{wrons_subl}.
 
  Similarly to the zero flow case, by looking at the series expansions of the relevant wavenumbers, we can make useful predictions about the transmission coefficients in this $\Lambda \gg 1$ regime. We start the analysis by solving~\eqref{disp_flow} in the asymptotic region $x \rightarrow +\infty$ and expressing  the obtained transmitted wavenumbers as power series in $\Lambda$,
\begin{align}
k_{t_1}&=\frac{\omega-e\Phi_0}{1+W_0}+\frac{1}{2}\frac{(\omega-e\Phi_0)^3}{(1+W_0)^4}\Lambda^{-2}+\mathcal{O}(\Lambda^{-4}),\\
k_{t_{2,3}} &= \pm\Lambda\sqrt{1-W_0^2}+W_0\frac{\omega-e\Phi_0}{1-W_0^2}+\mathcal{O}\left(\Lambda^{-1}\right). \label{k_prob}
\end{align}
The series expansions of the incident wavenumber $k_{in}$ and of the reflected wavenumbers, $k_{r_1}$ and $k_{r_2}$, are given, as previously, by~\eqref{momenta_exp}. From these $\Lambda$ expansions, we note that, unless $W_0 \sim 1$, we have $k_{t_{2,3}}, k_{r_2} \sim \mathcal{O}(\Lambda)$ and $k_{in},k_{r_1},k_{t_1} \sim \mathcal{O}(\Lambda^0)$. However, unlike in the zero-flow case, the difference between the high momentum transmitted modes $k_{t_{2,3}}$ and the high momentum reflected modes $k_{r_2}$ is not negligible for $\Lambda \gg 1$, being $\mathcal{O}\left(\Lambda\right)$. Hence, the appearance of such modes requires a large momentum change in our system. Therefore, if the potential is sufficiently smooth, we expect the conversion of incident modes $k_{in}$ into transmitted modes $k_{t_2}$ and $k_{t_3}$ and into reflected modes $k_{r_2}$ to be disfavoured in comparison with the low momentum transmission/reflection channel involving $k_{t_1}$ and $k_{r_1}$. In other words, we expect that the Wronskian condition \eqref{wrons_subl} will include only the low momentum channel, i.e.
\be
|R_1|^2= 1-\frac{\omega-e\Phi_0}{\omega}|T_1|^2. \label{consv1}
\ee

From this relation for the reflection coefficient, we would conclude that superradiance occurs for all frequencies in region II ($\omega_{\text{crit}}<\omega<e\Phi_0$) when $\Lambda\gg 1$. Note that this conclusion certainly does not hold in the case of a general dispersive parameter. If the condition $\Lambda \gg 1$ is not satisfied, there can be a mixture between the high and low momentum channels. Consequently, as discussed before, the total reflection coefficient given by~\eqref{wrons_subl} is not necessarily larger than one. In such a case, a definite answer about superradiance can only be obtained by solving the differential equations at every spatial point $x$.    

Having analyzed the case of $W_0<1$, let us now fix $W_0>1$ and assume $0 < \omega< e\Phi_0$. As discussed previously, there are two transmission and two reflection channels available when $W_0>1$ and the scattering problem is now described by~\eqref{sol_model3}. Since we are interested in the regime of large $\Lambda$, we calculate the wavenumber of the transmitted modes up to next-to-leading order terms,
\begin{align}
k_{t_1}&=\frac{\omega-e\Phi_0}{1+W_0}+\frac{1}{2}\frac{(\omega-e\Phi_0)^3}{(1+W_0)^4}\Lambda^{-2}+\mathcal{O}(\Lambda^{-4}), \notag \\
k_{t_2}&=\frac{\omega-e\Phi_0}{W_0-1}-\frac{1}{2}\frac{(\omega-e\Phi_0)^3}{(W_0-1)^4}\Lambda^{-2}+\mathcal{O}(\Lambda^{-4}). \label{asympkts}
\end{align}
 By direct substitution of these expressions into~\eqref{wrons_subl}, one can straightforwadly determine the reflection coefficient in powers of $\Lambda$, 
\be \label{wrons_subl2}
|R_1|^2+\left|\frac{k_{r_2}}{k_{in}}\right||R_2|^2= 1 -\frac{\omega - e \Phi_0}{ \omega } \left(\left| T_1 \right|^2 -  \left| T_2 \right|^2 \right),
\ee
plus terms of $\mathcal{O} \left(\Lambda ^{-2} \right)$. Note that this second channel $T_2$ is present even in the absence of dispersion, being an upstream mode which is swept downstream by a superluminal flow. From the equation above, it also becomes evident that the relation between the norms $|T_1|$ and $|T_2|$ of the two transmission channels determines the occurrence or not of superradiance.

\line

\noindent\tbf{Critical case:} An interesting situation to be analyzed is the critical case $\omega=\omega_{\text{crit}}$, which corresponds to the boundary between regions I and II in FIG.~\ref{regions_lab_sub}. In this scenario, the background flow $W_0$, when expanded in powers of $\Lambda$, relates to the critical frequency according to the following expression,
\be \label{ineq_w0}
W_0 = 1 - \frac{3}{2}\left(\omega_{\text{crit}}-e\Phi_0\right)^{\frac{2}{3}}\Lambda^{-\frac{2}{3}} +\mathcal{O}\left(\Lambda^{-\frac{4}{3}}\right).
\ee
%
Note that the previous analysis leading to~\eqref{consv1} relied on series expansions (see~\eqref{k_prob}) which are not valid when $W_0-1 \sim \mathcal{O}\left(\Lambda^{-\frac{2}{3}} \right)$. Therefore, in order to analyze the possibility of superradiance in the critical case, we cannot use~\eqref{consv1}; instead, we have to start from the original Wronskian relation~\eqref{wrons_subl}. 
\begin{figure}
\begin{center}
\includegraphics[scale=1.0]{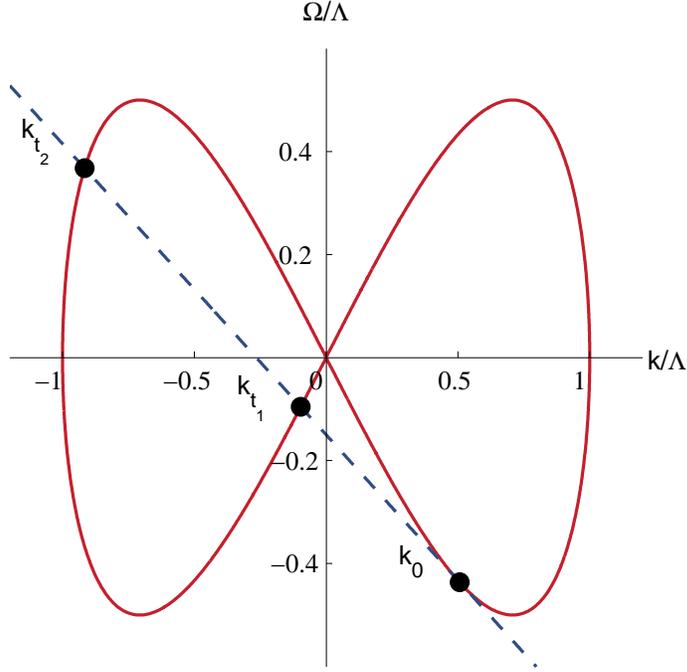}\caption{The subluminal dispersion relation in the fluid frame at the critical frequency. The blue dashed line represents the effective frequency $\Omega=\omega_{crit} - e \Phi_0 - kW_0$ at $x\rightarrow+\infty$.} 
\label{dispW0}
\end{center}
\end{figure}
In the critical regime, the dispersion relation (see FIG.~\ref{dispW0}) has three distinct solutions. Two of these solutions, denoted by $k_{t_1}$ and $k_{t_2}$, have positive group velocities in the lab frame and, therefore, are identified as transmitted modes. The other solution, denoted by $k_0$, is a degenerate double root and, consequently, has a vanishing group velocity in the lab frame. In order to obtain the reflection coefficient for the scattering problem, we first expand $k_{t_1}$ and $k_{t_2}$ as power series in $\Lambda$,
\begin{align}
& k_{t1} = \frac{\omega - e \Phi_0}{2} + \frac{3}{8}\left(\omega - e \Phi_0 \right)^{\frac{5}{3}}\Lambda^{-\frac{2}{3}}   + \mathcal{O}\left(\Lambda ^{-\frac{4}{3}}\right), \label{kt1fin}\\
 &k_{t2} = -2 \left(\omega - e \Phi_0 \right)^{\frac{1}{3}}\Lambda^{\frac{2}{3}} +\mathcal{O}(1) \notag,
\end{align}   
and then substitute the obtained expressions into~\eqref{wrons_subl}. The final result is given by 
\be \label{wrons_subl22}
|R_1|^2+\left|\frac{k_{r_2}}{k_{in}}\right||R_2|^2 = 1 -\frac{\omega - e \Phi_0}{ \omega } \left(\left| T_1 \right|^2 - 3 \left| T_2 \right|^2 \right),
\ee
plus terms of order $\mathcal{O} \left( \Lambda ^{-\frac{2}{3}} \right)$. Observe again the importance of the relative sign of the norm in the two transmission channels.
Note that the scattering process converts incident modes with wavenumber $k_{in} \approx \omega + \mathcal{O}(\Lambda^{-1})$ into transmitted modes of large wavenumber $k_{t_2}\approx \mathcal{O}(\Lambda ^{\frac{2}{3}})$, which can only be balanced by the high momentum reflected modes $k_{r_2} \approx \mathcal{O}\left(\Lambda\right)$. Another possibility is the conversion of the incident modes $k_{in} \approx \omega + \mathcal{O}(\Lambda^{-1})$ into transmitted modes of wavenumber $k_{t_1} \approx (\omega - e\Phi_0)/2 $, which is comparable to the low momentum $k_{r_1}$ channel. Because of the high momentum change required by the second channel (involving $k_{r_2}$ and $k_{t_2}$), we expect the first channel (involving $k_{r_1}$ and $k_{t_1}$) to be favoured in our scattering experiment.
Since $\omega_{\text{crit}} < e \Phi_0$, we therefore deduce that the RHS of~\eqref{wrons_subl22} is always greater than one in the limit $\Lambda \gg 1$. In other words, low-frequency waves in the critical regime are superradiantly scattered in our toy-model if small subluminal corrections are added to the dispersion relation.

\subsection{Superluminal dispersion}  \label{superlumsec}

We now turn to superluminal scattering processes, which can be quite different compared to subluminal ones since there does not exist any notion of a horizon or mode-independent blocking region for high momentum incident modes (the group velocity $d\omega/dk$ is unbounded as a function of $k$ and only the low frequency modes which possess quasilinear dispersion experience a blocking region in such flows). We will follow the standard treatment \cite{coutant} of analogue black holes with superluminal dispersion and analyze the transmission of an incoming wave from $-\infty$ through to $+\infty$. 

\begin{figure}
\begin{center}
\includegraphics[scale=1.0]{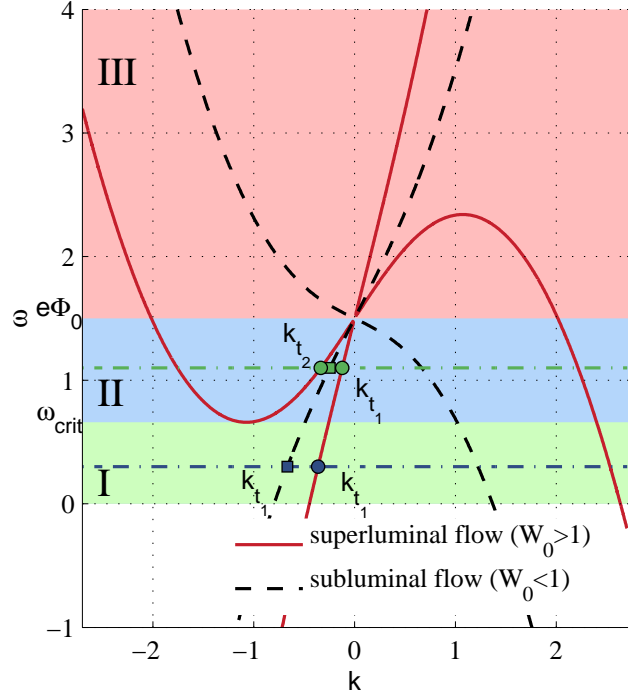}\caption{The superluminal dispersion curve in the lab frame ($\Lambda=1$, $e\Phi_0=1.5$) at $x\rightarrow + \infty$ for two different flow velocities: $W_0>1$ (red solid curve) and $W_0<1$ (black dashed curve). The intervals of frequency indicated refer to the red solid curve: region I (green, $0<\omega<\omega_\text{crit}$), region II  (blue, $\omega_\text{crit}<\omega<e\Phi_0$) and region III (light red, non-superradiant region). The intersections of the horizontal lines with the red solid and the black dashed curves indicate transmitted modes at that frequency. Note that, when $x\rightarrow -	\infty$, the dispersion is described by the green dash-dotted curve of FIG.~\ref{superlum}.
 \label{regions_final}} 
\end{center}
\end{figure}

The relevant dispersion relation in the superluminal case is depicted in FIG.~\ref{regions_final}. Given a superluminal flow $W_0>1$ (solid red curve in FIG.~\ref{regions_final}), there exists an interval of frequencies $0<\omega<\omega_\text{crit}$ (region I in FIG.~\ref{regions_final}) for which only two propagating modes are admitted, one right-moving and one left-moving in the lab frame.  For $\omega_\text{crit}<\omega<e\Phi_0$ (region II in FIG.~\ref{regions_final}), however, there are four propagating modes, two transmitted right-movers and two left-movers. 
The third possibility is a subluminal flow $W_0<1$ with $0 < \omega < e \Phi_0$, for which there are always two propagating modes (see the black dashed curve in FIG.~\ref{regions_final}).

Let us discuss first the cases in which only two propagating modes are available in the asymptotic limit $x\rightarrow+\infty$. Hence, the scattering is produced by an incident wave from $-\infty$ whose frequency $\omega$ satisfies $0<\omega < \omega_{\text{crit}}$ (if $W_0>1$) or $0< \omega < e\Phi_0$ (if $W_0 < 1$). Note that exactly one of these two propagating modes is a left-moving mode. Imposing the boundary condition that no incoming mode is allowed at $+\infty$, we obtain the solution of~\eqref{E:model} corresponding to the scattering problem, 
\be \label{sol_model4}
f \rightarrow \begin{cases} \text{e}^{ik_{in} x} + R\text{e}^{ik_{r} x}, & x\rightarrow -\infty, \\  T_1\text{e}^{ik_{t_1} x}, & x\rightarrow \infty, \end{cases}
\ee
plus exponentially decaying channels which do not contribute directly to the generalized Wronskian current calculated at $x\rightarrow \pm \infty$. Note that the wavenumbers $k_{in}$ and $k_{r}$ are the same ones that appear in~\eqref{momenta2} for the superluminal $W=0$ case and the wavenumber $k_{t}$ represents the only available transmission channel.

On the other hand, if $W_0>1$ and the frequency $\omega$ of the incident wave satisfies $\omega_{\text{crit}}< \omega < e\Phi_0$ (region II in FIG.~\ref{regions_final}), then there are, in principle, two extra propagating channels available (four in total, as discussed above). However, because of the boundary condition imposed at $x\rightarrow +\infty$, only one extra transmission channel has to be considered (the other extra channel is always left-moving at $x \rightarrow + \infty$). The scattering solution is then given by 
\be \label{sollmodel4b}
f \rightarrow \begin{cases} \text{e}^{ik_{in} x} + R\text{e}^{ik_{r} x}, & x\rightarrow -\infty, \\  T_1\text{e}^{ik_{t_1} x}+T_2\text{e}^{ik_{t_2} x}, & x\rightarrow \infty, \end{cases}
\ee 
where $k_{in}$ and $k_{r}$ are again given by~\eqref{momenta2} and $k_{t_1}$ and $k_{t_2}$ are the wavenumbers of the transmitted modes. Note that we have once again omitted the exponential decaying mode at $x\rightarrow -\infty$ since it does not affect directly the generalized Wronskian.

Using~\eqref{action} to evaluate the functional $X$ in both asymptotic regions, we obtain, similarly to the subluminal case, the following relation between the reflection and transmission coefficients,
%
%
\be \label{wrons_supl}
|R|^2 = 1 - \frac{\Lambda}{\sqrt{\Lambda ^2 + 4 \omega ^2}} \left(\sum_n \frac{v_{g_n}}{k_{in}} \left( \omega - e\Phi_0 - k_{t_n} W_0 \right)  \left| T_n \right|^2   \right),
\ee
where the sum is over one or two transmission channels, depending on $\omega$ and whether $W_0>1$ or $W_0<1$. 
Here, $v_{g_1}$ and $v_{g_2}$ are the group velocities of the transmitted modes $k_{t_1}$ and $k_{t_2}$, which are always positive by definition. Furthermore, it is possible to show, for frequencies $0<\omega<e \Phi_0$, that the effective frequency $\Omega = \omega - e\Phi_0 - k_{t_n} W_0$ is always negative for $k_{t_1}$ modes and always positive for $k_{t_2}$ modes. Therefore, since only the $n=1$ transmission channel is available for frequencies lying in region I of FIG.~\ref{regions_final}, we conclude that the RHS of~\eqref{wrons_supl} is greater than 1 and, therefore, the scattering is always superradiant. 

The situation for $W_0<1$ and $0 < \omega < e\Phi_0 $ is similar: only the first transmission channel is available and superradiance always occurs. However, for frequencies located in region II, we cannot so easily conclude superradiance since the extra transmission channel $k_{t_2}$ contributes an overall negative factor in~\eqref{wrons_supl}. To obtain a conclusive answer, one would need to know the detailed structure of $W(x)$ and $V(x)$ in the intermediate regime and solve the equations not only in the asymptotic regions but at every point $x$.    

\line 

\noindent\tbf{Large $\Lambda$ approximation:} In order to better understand the scattering of an incident wave whose frequency is located in region II of FIG.~\ref{regions_final}, we shall consider small deviations from the non-dispersive limit, i.e.~$\Lambda \gg 1$. In such a case, we can expand the two transmission channels, $k_{t_1}$ and $k_{t_2}$, in powers of $\Lambda$,
\begin{align}
k_{t_1}&=\frac{\omega-e\Phi_0}{1+W_0}-\frac{1}{2}\frac{(\omega-e\Phi_0)^3}{(1+W_0)^4}\Lambda^{-2}+\mathcal{O}(\Lambda^{-4}),\\
k_{t_2}&=\frac{\omega-e\Phi_0}{W_0-1}+\frac{1}{2}\frac{(\omega-e\Phi_0)^3}{(W_0-1)^4}\Lambda^{-2}+\mathcal{O}(\Lambda^{-4}), \label{kt2spl}
\end{align}
and substitute the obtained wavenumbers into~\eqref{wrons_supl} in order to determine the reflection coefficient for the scattering, 
\be \label{wrons_supl2}
|R|^2 = 1 -\frac{\omega - e \Phi_0}{ \omega } \left(\left| T_1 \right|^2 -  \left| T_2 \right|^2 \right) +\mathcal{O} \left(1/ \Lambda ^{2} \right).
\ee
Since we assume $\omega < e \Phi_0$ in region II, this reflection coefficient is larger than 1 whenever $|T_2|<|T_1|$. As explained above, whether this condition is satisfied or not in a general model would depend on the detailed structure of $W(x)$ and $V(x)$ in the intermediate regime \cite{coutant}. From a practical point of view, in order to maximize the potential for superradiance in an experiment with a superluminally dispersive medium, one should choose the asymptotic flow $W_0$ as small as possible while still being superluminal ($W_0 > 1$) as this would minimize the size of region II and the extra positive-effective-frequency transmission channels therein.

Choosing the flow as such to maximize the $T_1$ channel is also consistent with our intuition that scattering favors the channel which most closely matches the momentum of the reflected mode; in this case the wavenumber $k_{t_1}$ is closer to $k_{r}$ than $k_{t_2}$ is. This prediction is confirmed in the step function model for which $V(x)=e\Phi_0\Theta(x)$ and $W(x)=W_0 \Theta(x)$. In such a case, one can impose the appropriate boundary conditions at $x=0$ discussed in the appendix to obtain the following reflection and transmission coefficients,
\begin{align}
R = \frac{e\Phi_0 + \omega (W_0 - 1)}{ \omega (W_0 + 1) - e\Phi_0} + \mathcal{O}\left( \Lambda ^ {-1}\right) ,\\
T_1= \frac{ \omega (W_0 + 1)}{ \omega (W_0 + 1) - e\Phi_0} + \mathcal{O}\left( \Lambda ^ {-1}\right) ,\\
T_2= \frac{ \omega (W_0 - 1)}{ \omega (W_0 + 1) - e\Phi_0} + \mathcal{O}\left( \Lambda ^ {-1}\right).
\end{align} 
We can also show that the coefficient correponding to the omitted exponential decaying mode in~\eqref{sollmodel4b} is of order $\mathcal{O}\left( \Lambda ^ {-2}\right)$. 

Comparing $T_1$ and $T_2$ above and using the fact that $W_0>1$, one can see that $|T_1|>|T_2|$ at zeroth order in $\Lambda$, which implies superradiance and confirms our expectations. Alternatively, one can verify the occurence of superradiance by directly analyzing the reflection coefficient $R$ above. It is straightforward to see that $|R^2|>1$ at lowest order in $\Lambda$.   

\line

\begin{figure}
\begin{center}
\includegraphics[scale=1.0]{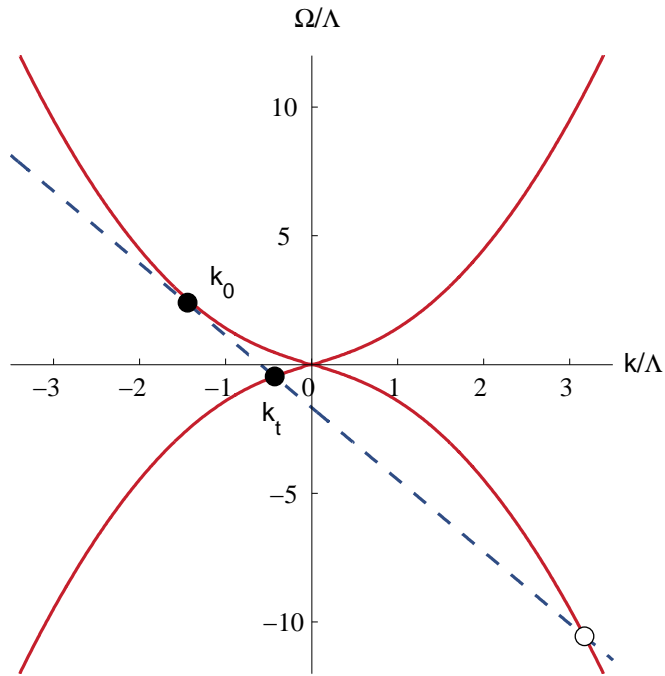}\caption{The superluminal dispersion relation in the fluid frame  at the critical frequency. The blue dashed line represents the effective frequency $\Omega=\omega_{crit} - e \Phi_0 - kW_0$ at $x\rightarrow+\infty$.  
\label{dispW02}}
\end{center}
\end{figure}

\noindent\tbf{Critical case:} Another interesting possibility that we now consider in detail is the critical regime $\omega=\omega_{\text{crit}}$. This situation corresponds to the boundary between regions I and II in FIG.~\ref{regions_final} and is depicted, in the fluid frame, in FIG.~\ref{dispW02}. The relation between the background flow $W_0$ and the critical frequency $\omega_{\text{crit}}$ is given by the following expression,
\be
W_0 = 1 + \frac{3}{2}\left(\omega_{\text{crit}} - e \Phi_0 \right)^{\frac{2}{3}}\Lambda^{-\frac{2}{3}} + \mathcal{O}\left(\Lambda ^{-\frac{4}{3}}\right).
\ee
Note that, since $W_0-1 \sim \mathcal{O}\left(\Lambda^{-\frac{2}{3}} \right)$, the $\Lambda$ expansion~\eqref{wrons_supl2} of the generalized Wronskian obtained previously is not valid in the present case (check the denominators in~\eqref{kt2spl}). Consequently, we shall need different $\Lambda$ expansions in order to obtain an appropriate expression for the reflection coefficient. 

Like in the critical subluminal case, the dispersion relation has three distinct roots: the double root $k_0$ (with vanishing group velocity in the lab frame), a right-moving mode (with negative group velocity in the lab frame) and a transmitted mode (with positive group velocity in the lab frame) whose wavenumber $k_{t}$ is given by 
\begin{align}
 k_{t} = \frac{\omega - e \Phi_0}{2} - \frac{3}{8}\left(\omega - e \Phi_0 \right)^{\frac{5}{3}}\Lambda^{-\frac{2}{3}}   + \mathcal{O}\left(\Lambda ^{-\frac{4}{3}}\right). \label{ktkrfin}
\end{align}   
Applying as a boundary condition the fact that only right moving modes are allowed at $+\infty$, we obtain, after substituting the relevant quantities into~\eqref{wrons_supl}, the relation between the reflection and transmission coefficients for the scattering process,
\be \label{wrons_supl3}
|R|^2 = 1 -\frac{\omega - e \Phi_0}{ \omega } \left| T \right|^2 + \mathcal{O} \left( \Lambda ^{-\frac{2}{3}} \right).
\ee
Since $\omega_{\text{crit}} < e \Phi_0$, we conclude that the RHS of the equation above is always greater than one when $\Lambda \gg 1$. In summary, superradiance is expected to occur in the superluminal critical case for small deviations from the non-dispersive regime.

\subsection{Inertial motion superradiance}

Throughout this paper, inspired by the usual condition for rotational superradiance in the black hole case, i.e.~$\omega - m\Omega_h<0$, we analyzed only scattering problems in which the frequency of the incident mode satisfies $\omega - e \Phi_0 < 0$. However, since it is the effective frequency $\Omega$ that appears in~\eqref{action}, we conclude that the amplification of an incident mode can also occur for $\omega -e\Phi_0>0$ given that $\omega - e \Phi_0 - kW_0 < 0$. In particular, even when $\Phi_0 = 0$ superradiant scattering will be possible. However, being due exclusively to inertial motion in the system, this kind of superradiance is outside the scope of our work. In fact, inertial motion superradiance has long been known in the literature as the anomalous Doppler effect and the condition for negative effective-frequency modes is referred to as the Ginzburg-Frank condition~\cite{ginzburg1}. Several phenomena in physics, like the Vavilov-Cherenkov effect and the Mach cones (which appear in supersonic airplanes) can be understood in terms of inertial motion superradiance~\cite{super1}. For a detailed analysis of inertial motion superradiance, we refer the reader to~\cite{super1}.

\section{Applications: axisymmetric systems} \label{applications}
Having analyzed superradiance in simple 1+1-dimensional toy models with modified dispersion relations, we will now discuss how the ideas presented in this paper can be generalized to more realistic situations based on analogue models of gravity. Our starting point is a general 2+1-dimensional, axisymmetric and irrotational fluid flow with background velocity $\mathbf{v}$ given by
\be
\mathbf{v} \equiv v_r(r)\hat{\mathbf{r}} + v_{\phi} (r) \boldsymbol{\hat{\boldmath{\phi}}} = -\frac{A}{r} \hat{\mathbf{r}} + \frac{B}{r} \boldsymbol{\hat{\boldmath{\phi}}},
\ee
where $A$ and $B$ are constants and $(r,\phi)$ are the usual polar coordinates. Velocity perturbations $\delta \mathbf{v}$ of the background flow can be conveniently described by a scalar field $\psi$, which relates to $\delta \mathbf{v}$ through equation~$\delta \mathbf{v} = \nabla \psi$. We denote the propagation speed of these perturbations by $c$. The idea of analogue gravity is derived from the observation that the differential equation satisfied by the perturbations $\psi$, i.e.
\be \label{usual_wave}
-\left(\partial_t + \nabla \cdot \mathbf{v} \right)\left(\partial_t + \mathbf{v} \cdot \nabla \right) \psi + c^2 \nabla ^ 2 \psi = 0,
\ee
can be cast into a Klein--Gordon equation in an effectively curved spacetime geometry. This connection between hydrodynamics and gravity is responsible for many important results, see~\cite{review} for a detailed review. 

One of the successes of Unruh's~\cite{unruh} original idea of using sound waves to study gravitational phenomena is that it can be extended to many other physical systems, like gravity waves in open channel flows~\cite{Unruh_ralf} and density perturbations in Bose-Einstein condensates~\cite{garay}. An important feature of such systems is that~\eqref{usual_wave} is only accurate in certain regimes; at sufficiently small distance scales (e.g.~wavelengths comparable to the fluid depth in open channels), the dispersion relation is not linear anymore and~\eqref{usual_wave} has to be replaced by\footnote{Note that this equation includes only fourth order corrections. A full description of the system will possibly include also higher order terms.}
\be \label{mod_wave}
-\left(\partial_t + \nabla \cdot \mathbf{v} \right)\left(\partial_t + \mathbf{v} \cdot \nabla \right) \psi + c^2 \nabla ^ 2 \psi = \mp\frac{1}{\Lambda ^ 2} \nabla ^ 4 \psi,
\ee
where $\Lambda$ is a dispersive parameter and the upper (lower) sign corresponds to subluminal (superluminal) dispersion.

Even though superradiance has been subjected to extensive studies in the linear regime of analogue models, it has never been analysed before in the context of modified dispersion relations, as opposed to Hawking radiation (see e.g.~\cite{review} and references therein). The remarkable fact about the toy models introduced in this paper is that they can be used to analyze superradiance in realistic dispersive analogue models of gravity satisfying~\eqref{mod_wave}.

Firstly, the electromagnetic interaction term $e\Phi_0$ appearing in our toy models is analogous to the rotational term $m\Omega$ in axisymmetric analogue models, where $m$ is the azimuthal number and $\Omega$ is the angular velocity. It is interesting to note that this duality is manifest in real black holes: both electromagnetic~\cite{bekenstein} and rotational~\cite{misner} superradiance are possible.
Another essential ingredient for the occurrence of superradiance in analogue models of gravity is the presence of an event horizon, which allows no mode to escape from inside the analogue black hole. In our toy model, such behaviour is mimicked by an appropriate boundary condition imposed in the asymptotic limit $x \rightarrow \infty$. 
              
It is also important to address the usefulness of the generalized Wronskian~\eqref{nonzero_wronsk} in the context of axisymmetric systems. More precisely, we are going to show that, if all derivatives with respect to $x$ are replaced by derivatives with respect to $r$, then the Wronskian~\eqref{nonzero_wronsk}, when applied to solutions of~\eqref{mod_wave}, is independent of $r$. Indeed, by applying the \textit{ansatz} $\psi=\left(H(r)/\sqrt{r}\right) \, \text{e}^{i m \phi}\text{e}^{-i \omega t}$, we are able to separate~\eqref{mod_wave} and are left with a radial equation for $H$, 
\be \label{hequation}
H''''(r)+\alpha(r) H''(r)+\beta(r)H'(r)+\gamma(r)H=0,
\ee
where the coefficients $\alpha(r)$, $\beta(r)$ and $\gamma (r)$ are given by
\begin{align}
\alpha &= \frac{1-4m^2}{2r^2} \pm \Lambda^2(c^2-v_r^2), \notag \\
\beta &=\frac{4m^2 - 1}{r^3} \pm \Lambda^2\left(P(r) - \frac{c^2 - v_r^2}{r}\right), \label{alpbetgam} \\
\gamma &=\frac{25}{16r^4} - \frac{13m^2}{2r^4} + \frac{m^4}{r^4} \notag \\ &\pm \Lambda^2 \left(\frac{3}{4}\frac{c^2 - v_r^2}{r^2} - \frac{P(r)}{2r} + Q(r) \right). \notag
\end{align}
The functions $P(r)$ and $Q(r)$ appearing in the coefficients above are, up to a factor $\left(c^2-v_r^2\right)$, the same functions $P$ and $Q$ defined in Refs.~\cite{basak2,berti}. They can be expressed as 
\begin{align}
P &= \frac{c^2}{r}\frac{d}{dr}\left[\frac{r}{c^2}(c^2 - v_r^2) \right] + 2i v_r \left(\omega - \frac{mB}{r^2}\right), \notag \\
Q &= \left( \omega - \frac{mB}{r^2} \right)^2 - \frac{ m^2 c^2}{r^2} \label{PeQ} \\
  & + i \frac{c^2}{r}\frac{d}{dr} \left[\frac{rv_r}{c^2} \left( \omega - \frac{mB}{r^2} \right)\right], \notag
\end{align}
where the upper (lower) sign corresponds to subluminal (superluminal) dispersion. Finally, we note that~\eqref{hequation} is exactly the same as~\eqref{nonzeroflow} and, more remarkably, that the coefficients $\alpha$, $\beta$, $\gamma$ above satisfy conditions \eqref{wcond1} and \eqref{wcond2}. Consequently, the generalized Wronskian defined in~\eqref{nonzero_wronsk} can also be used in the context of axisymmetric analogue models of gravity, thus completing the connection between our toy models and realistic physical systems.

\section{Summary and final remarks}

We have proposed idealized systems to investigate multi superradiant scattering processes that are applicable to sub and superluminal dispersive fields. Perhaps the most important theoretical result obtained is related to the simplicity of the analytic expression for the particle number currents $J_n$ of the scattering channels $n$ (see~\eqref{action}), 
\begin{equation} \label{Eq:J}
J_n = \vert A_n \vert^2 \left.  \Omega \right\vert_{\pm \infty}  \left. \frac{d\omega}{dk} \right\vert_{k_n}.
\end{equation}
Notice that the expression above depends only on the amplitude, group velocity and effective frequency of the particular scattering channel. Moreover, this result is universal to all scattering processes discussed in this paper. Note also that, in principle, we could have normalized the modes so that the group velocity in \eqref{Eq:J} is absorbed into the coefficients $A_n$. Doing that would make the conservation equations (see e.g.~\eqref{tot_ref}) look much simpler. However, their dependence on the dispersive parameter $\Lambda$ would then also be hidden in these new coefficients.

\begin{figure}[h!]
\begin{center}
\includegraphics[scale=0.6]{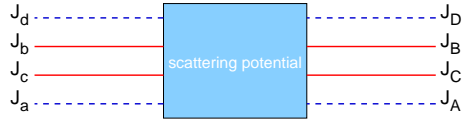}
\caption{The scattering processes for sub and superluminal dispersive fields, with two low (red solid) and two high (blue dashed) momentum degrees on each side of the potential, are represented by a general scattering process whose particle number currents are given by~(\ref{Eq:J}). \label{FIG:Summary}}
\end{center}
\end{figure}

\begin{figure}[h!]
\begin{center}
\includegraphics[scale=0.6]{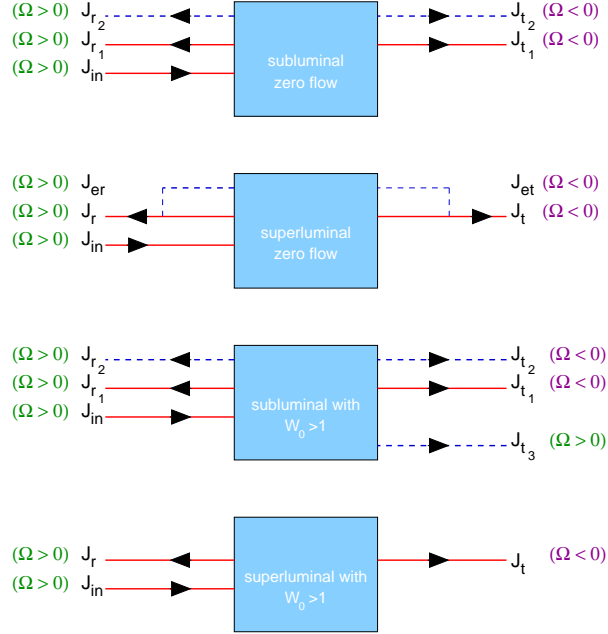}
\caption{Schematic representation of some of the scattering processes investigated in the paper. The arrows indicate the group velocity propagation. The particle number current direction is related to the relative sign between this group velocity and the effective frequency $\Omega$. This information, together with~\eqref{Jsum}, can help determine if superradiance occurs or not.  \label{FIG:Summary2}}
\end{center}
\end{figure}

Our findings link to standard scattering processes, allowing a deeper insight into superradiance. Let us consider scattering of up to four incident modes by a general scattering potential. There are up to four channels to the left $\{a,b,c,d\}$ and four to the right $\{A,B,C,D\}$ of the scattering potential. As a lesson from the analysis carried out before, one needs to be careful when assigning the propagation direction of each mode since its group velocity is dependent on the particular scattering potential and type of dispersion, see FIG.~\ref{FIG:Summary}. Furthermore, as long as the scattering potential is real, the total current to the left equals the total current to the right of the potential,
\begin{equation} \label{Jsum}
J_a + J_b + J_c + J_d = J_A + J_B + J_C + J_D \,.
\end{equation} 
The scattering potential is amplifying classical and quantum field excitations if, at the left side of the potential, the total outflux (i.e.~the reflected current) is larger then the total in-flux, $J^\mathrm{ref}_\mathrm{total}>J^\mathrm{in}_\mathrm{total}$. There are several scattering coefficients that can be considered: the reflection coefficient in each individual scattering channel,
\begin{equation}
P_{n} = \frac{J^\mathrm{ref}_{n}}{J^\mathrm{in}_\mathrm{total}},
\end{equation}
and the total reflection coefficient,
\begin{equation}
P_\mathrm{total} = \frac{J^\mathrm{ref}_\mathrm{total}}{J^\mathrm{in}_\mathrm{total}} = \frac{\sum_n J^\mathrm{ref}_{n}}{J^\mathrm{in}_\mathrm{total}} .
\end{equation}
A sufficient, but not necessary, condition for superradiance, is to demand no influx from the right and to strictly require negative effective frequency for the remaining channels to the right ---these conditions arise naturally at the event horizon of a rotating black hole. A surprising result of our analysis is that the presence of extra scattering channels can enhance the amplification effect. In FIG.~\ref{FIG:Summary2}, we illustrate the scattering diagrams for some of the examples discussed in the paper.

In addition, the present work can be related to recent experimental realizations of analogue black holes, some of which have even studied the Hawking emission process. In particular, \cite{weinfurtner} exhibits the first detection of the classical analogue of Hawking radiation using dispersive gravity waves in an open channel flow. 
One of the most important lessons to be learned from that work is that even though vorticity and viscosity effects cannot be completely removed from the experimental setup, they can be made extremely small. In fact, they can be reduced to the point that the results predicted by the irrotational and inviscid theory match the results obtained experimentally with considerable accuracy. Based on this fact, together with our present results, one might ask whether superradiance occurs (and can be observed) in the laboratory using dispersive gravity waves. The connection between such system and our analysis can be seen directly from the full dispersion relation for gravity waves in a fluid of constant depth $h$~\cite{lamb}, 
\begin{align}
\omega ^ 2(k) &= \left(gk + \frac{\sigma}{\rho}k^3 \right) \tanh \left( kh\right) \notag \\
  &= gh k ^ 2 -\left( \frac{gh^ 3}{3} - \frac{\sigma h}{\rho} \right)k^ 4 + \mathcal{O}\left(k^6 \right),
\end{align}
where $g$ is the gravitational acceleration and $\sigma$ and $\rho$ correspond, respectively, to the surface tension and the density of the fluid. If first-order deviations from the shallow water limit ($kh \ll 1$) are considered, we recover the quartic dispersion relation analyzed in this paper.

In other words, non-shallow gravity waves impinging on a rotating analogue black hole (e.g.~a draining 'bathtub' vortex) satisfy~\eqref{mod_wave} with $c^2 = gh$ and a dispersive parameter $\Lambda^{-2} = \left|gh^3/3 - \sigma h / \rho \right|$. Based on the existent analysis of superradiance of linear fields in open channel flows~\cite{Unruh_ralf, berti} together with our discussion in Section~\ref{applications}, we expect superradiance to also be manifest for non-shallow gravity waves. 
This relation between dispersion and superradiance in open channels is currently being further investigated by the authors and will be the subject of a future work.

    Another class of analogue black holes, which was recently set up in the laboratory~\cite{bec_bh} and which might be used in the future to produce superradiant scattering processes, consists of Bose-Einstein condensates~\cite{garay, review}. The analogy with gravity arises when one considers the Gross-Pitaevskii equation and uses the Madelung representation of the condensate wave function. If the eikonal approximation is used and axisymmetry is assumed, it is possible to show that perturbations around a background condensate obey a superluminal dispersion relation given by
\be
\omega^2 = 4 \pi \hbar ^ 2 \frac{n_0 a}{m^2} k^2  + \left(\frac{\hbar}{2m}\right)^2 k^4,
\ee
where $m$ is the mass of a single boson, $a$ is the scattering length and $n_0(r)$ is the background density. Such perturbations are described by~\eqref{mod_wave} with $c^2 = 4 \pi \hbar ^ 2 n_0 a /m^2$ and $\Lambda = 2m/ \hbar$. 
Based on our work, the most obvious conclusion we can draw is to expect superradiance to be manifest also in low-frequency BEC scattering experiments. However, since the dispersion relation is superluminal, there is no notion of a mode independent blocking region (see section~\ref{superlumsec}) and, consequently, it is not clear if the appropriate boundary conditions will be sufficient to guarantee superradiance. Additionally, quantized vortices may be present in such systems, restricting the angular momentum to integer multiples of $\hbar$. The physics of these quantized vortices with respect to superradiance and instabilities is also unclear at this point and more investigation is needed to understand their role in possible BEC scattering processes.

\ack
MR was partially supported by FAPESP. SW was supported by Marie Curie Career Integration Grant (MULTI-QG-2011), the SISSA Young Researchers Grant (Black hole horizon effects in fluids and superfluids), and the Fqxi Mini grant (Physics without borders). We wish to thank Carlos Barcel\'o for his comments. SW would like to thank Matt Visser for stimulating discussions.

\appendix

\section*{Appendix} \label{BCS}

Throughout this paper we solve simple models based on step functions for the external potential $V(x) = e\Phi_0\Theta(x)$ and for the background velocity $W(x) = W_0\Theta(x)$, where $\Theta(x)$ is the Heaviside function. Since this function is characterized by a discontinuity at $x=0$, it is important to analyze what happens to the wavefunction, i.e.~the solution to~\eqref{EOM} or~\eqref{nonzeroflow}, at $x=0$. The same problem arises in 1D tunneling problems in quantum mechanics when the potential barrier in the Schr\"odinger equation is modelled by a step function. In such situations, one has to impose the continuity of the wavefunction and its first derivative at the discontinuity point in order to determine the reflection and the transmission coefficients.    

  Let us first consider~\eqref{EOM}, which is valid only for zero background flows. Following the standard procedure used in quantum mechanics (see e.g.~\cite{merzb}), we integrate the differential equation over a small region $(-\epsilon, + \epsilon)$ around the discontinuity $x=0$ and then take the limit of the obtained expression as $\epsilon \rightarrow 0$. Starting from~\eqref{EOM} and repeating this procedure three times, one can show that $f(x)$ and its derivatives up to third order are all continuous at $x=0$.      

    We also have to deal with the generalization to non-zero flows, given by~\eqref{nonzeroflow}. The situation now is more complicate since the differential equation involves derivatives of $\Theta(x)$ (i.e.~delta functions). Repeating the procedure described above, one can show that the function $f$ and its first order derivative are still continuous at $x=0$. The continuity of the second and third order derivatives, because of the delta functions, now depends on $\Theta(0)$, i.e.~it depends on the choice of the flow velocity and of the external potential exactly at $x=0$. Using the half-maximum convention of $\Theta(0)=1/2$, we can further prove that the second order derivative $f''(x)$ is also continuous at $x=0$ and that the third order derivative satisfies the following relation,

\begin{align}
 f'''(0^+) &- f'''(0^-) = \Lambda^2  W_0^2 f'(0)  \\
& \hspace{11mm}   - i \Lambda^2  W_0(\omega - e \Phi_0) f(0). \nonumber
\end{align}    

\section*{References}

\begin{thebibliography}{10}
\expandafter\ifx\csname url\endcsname\relax
  \def\url#1{{\tt #1}}\fi
\expandafter\ifx\csname urlprefix\endcsname\relax\def\urlprefix{URL }\fi
\providecommand{\eprint}[2][]{\url{#2}}

\bibitem{unruh}
Unruh W~G 1981 {\em Phys. Rev. Lett.\/} {\bf 46} 1351--1353

\bibitem{review}
Barcel\'o C, Liberati S and Visser M 2011 {\em Living Reviews in Relativity\/}
  {\bf 14}

\bibitem{unruh_2}
Unruh W~G 2008 {\em Phil. Trans. R. Soc. A\/} {\bf 366} 2905--2913

\bibitem{philbin}
Philbin T~G, Kuklewicz C, Robertson S, Hill S, K\"onig F and Leonhardt U 2008
  {\em Science\/} {\bf 319} 1367--1370

\bibitem{rousseaux}
Rousseaux G, Mathis C, Maissa P, Philbin T~G and Leonhardt U 2008 {\em New J.
  Phys.\/} {\bf 10} 053015

\bibitem{bec_bh}
Lahav O, Itah A, Blumkin A, Gordon C, Rinott S, Zayats A and Steinhauer J 2010
  {\em Phys. Rev. Lett.\/} {\bf 105}(24) 240401

\bibitem{hid_jump}
Jannes G, Piquet R, Maissa P, Mathis C and Rousseaux G 2011 {\em Phys.Rev.\/}
  {\bf E83} 056312

\bibitem{belgiorno}
Belgiorno F, Cacciatori S~L, Clerici M, Gorini V, Ortenzi G, Rizzi L, Rubino E,
  Sala V~G and Faccio D 2010 {\em Phys. Rev. Lett.\/} {\bf 105} 203901

\bibitem{weinfurtner}
Weinfurtner S, Tedford E~W, Penrice M~C, Unruh W~G and Lawrence G~A 2011 {\em
  Phys. Rev. Lett.\/} {\bf 106} 021302

\bibitem{unruh_comment}
Sch{\"u}tzhold R and Unruh W~G 2011 {\em Phys. Rev. Lett.\/} {\bf 107}(14)
  149401

\bibitem{belgiorno_reply}
Belgiorno F, Cacciatori S~L, Clerici M, Gorini V, Ortenzi G, Rizzi L, Rubino E,
  Sala V~G and Faccio D 2011 {\em Phys. Rev. Lett.\/} {\bf 107}(14) 149402

\bibitem{ralf_1}
Sch{\"u}tzhold R 2011  (\textit{Preprint} \eprint{1110.6064})

\bibitem{angus_stefano}
Liberati S, Prain A and Visser M 2012 {\em Phys. Rev. D\/} {\bf 85}(8) 084014

\bibitem{unruh_3}
Unruh W~G 1995 {\em Phys. Rev. D\/} {\bf 51} 2827--2838

\bibitem{brout}
Brout R, Massar S, Parentani R and Spindel P 1995 {\em Phys. Rev. D\/} {\bf 52}
  4559--4568

\bibitem{jacobson}
Jacobson T 1991 {\em Phys. Rev. D\/} {\bf 44} 1731--1739

\bibitem{corinne}
Manogue C~A 1988 {\em Annals of Physics\/} {\bf 181} 261 -- 283 ISSN 0003-4916

\bibitem{super1}
Bekenstein J~D and Schiffer M 1998 {\em Phys. Rev. D\/} {\bf 58}(6) 064014

\bibitem{super2}
Richartz M, Weinfurtner S, Penner A~J and Unruh W~G 2009 {\em Phys. Rev. D\/}
  {\bf 80}(12) 124016

\bibitem{staro1}
{Starobinsky} A~A 1973 {\em Sov. Phys. JETP\/} {\bf 37} 28

\bibitem{staro2}
{Starobinsky} A~A and {Churilov} S~M 1974 {\em Sov. Phys. JETP\/} {\bf 38} 1--5

\bibitem{basak}
Basak S and Majumdar P 2003 {\em Class. Quant. Grav.\/} {\bf 20} 3907--3914

\bibitem{wald}
Wald R 1984 {\em General Relativity\/} (University of Chicago Press)

\bibitem{zeldovich}
{Zel'Dovich} Y~B 1972 {\em Sov. Phys. JETP\/} {\bf 35} 1085

\bibitem{misner}
Misner C~W 1972 {\em Bull. Am. Phys. Soc.\/} {\bf 17} 472

\bibitem{klein}
Klein O 1929 {\em Z. Phys.\/} {\bf 53} 157

\bibitem{Laing}
Diver D~A and Laing E~W 1990 {\em J. Phys. A: Math. Gen.\/} {\bf 23} 1699

\bibitem{relat1}
Bilic N 1999 {\em Class. Quant. Grav.\/} {\bf 16} 3953-3964

\bibitem{relat2}
Visser M and Molina-Paris C 2010 {\em New J.
  Phys.\/} {\bf 12} 095014

\bibitem{relat3}
Fagnocchi S, Finazzi S, Liberati S, Kormos M and Trombettoni A 2010 {\em New J.
  Phys.\/} {\bf 12} 095012

\bibitem{coutant}
Coutant A, Parentani R and Finazzi S 2012 {\em Phys. Rev. D\/} {\bf 85}(2)
  024021

\bibitem{ginzburg1}
Ginzburg V~L and Frank I~M 1947 {\em Dokl. Akad. Nauk SSSR\/} {\bf 56} 583

\bibitem{Unruh_ralf}
Sch{\"u}tzhold R and Unruh W~G 2002 {\em Phys. Rev. D\/} {\bf 66} 044019

\bibitem{garay}
Garay L~J, Anglin J~R, Cirac J~I and Zoller P 2000 {\em Phys. Rev. Lett.\/}
  {\bf 85} 4643--4647

\bibitem{bekenstein}
Bekenstein J~D 1973 {\em Phys. Rev. D\/} {\bf 7} 949--953

\bibitem{basak2}
Basak S and Majumdar P 2003 {\em Class. Quant. Grav.\/} {\bf 20} 2929--2936

\bibitem{berti}
Berti E, Cardoso V and Lemos J~P~S 2004 {\em Phys. Rev. D\/} {\bf 70} 124006

\bibitem{lamb}
Lamb H 1932 {\em Hydrodynamics\/} (Cambridge University Press)

\bibitem{merzb}
Merzbacher E 1998 {\em Quantum Mechanics\/} (Wiley)

\end{thebibliography}

\providecommand{\newblock}{}

\end{document}